\documentstyle[12pt,epsf]{article}
\def\Lc#1{{\rm Lc}_{#1}}
\def\Lcd#1{{\rm Lcd}_{#1}}
\def\Ld#1{{\rm Ld}_{#1}}
\def\Le#1{{\rm Le}_{#1}}
\def\Li2{{\rm Li_2}}
\def\P{{\cal P}}
\def\I{{\rm I}}
\def\Ihat{\hat{\rm I}}
\def\e{\epsilon}
\def\Gn{\Delta_n}
\def\Gnhat{\hat\Delta_n}
\def\G3{\Delta_3}
\def\G3hat{\hat\Delta_3}
\def\g{\gamma}
\def\ghat{\hat\gamma}
\def\chat{\hat {c}}

\begin{document}

\begin{titlepage}
\begin{flushright}
DTP/96/96\\
hep-ph/9612413\\
\end{flushright}
\vspace{1cm}
\begin{center}
{\Large\bf One-Loop Tensor Integrals in Dimensional Regularisation }\\
\vspace{1cm}
{\large
J.~M.~Campbell, E.~W.~N.~ Glover and
D.~J.~Miller\footnote{Address after 1 January 1997,
Rutherford Appleton Laboratory,  Chilton, Didcot, Oxon,
OX11 0QX, England}}\\
\vspace{0.5cm}
{\it
Physics Department, University of Durham,\\
Durham DH1~3LE,
England} \\
\vspace{0.5cm}
{\large December 1996}
\vspace{0.5cm}
\end{center}
\begin{abstract}
We show how to evaluate tensor one-loop integrals in momentum space
avoiding the usual plague of Gram determinants.  We do this by
constructing combinations of $n$- and $(n-1)$-point scalar integrals
that are finite in the limit of vanishing Gram determinant.
These non-trivial combinations of dilogarithms, logarithms and constants
are systematically obtained by either differentiating with respect
to the external parameters - essentially yielding scalar integrals
with Feynman parameters in the numerator - or by developing
the scalar integral in $D=6-2\e$ or higher dimensions.
An additional advantage is that other spurious kinematic singularities are also
controlled.
As an explicit example, we develop the tensor integrals and associated
scalar integral combinations for processes where the internal particles are
massless and where up to five (four massless and one massive)
external particles are involved.
For more general processes, we present the
equations needed for
deriving the relevant combinations of scalar integrals.
\end{abstract}
\end{titlepage}

\section{Introduction}
\setcounter{equation}{0}

One of the most important ingredients in the search for
``signals'' of new
phenomena in high energy particle physics experiments is a
precise knowledge of the expectations from standard physics;
the ``background''.
Usually this involves perturbative calculations of differential
cross sections within the standard
model. Many such radiative corrections have been carried out
and require the evaluation
of one-loop integrals which arise directly
from a Feynman diagrammatic approach.
Often these integrals need to be performed in
an arbitrary dimension in order to isolate any infrared
and ultraviolet divergences that may be present \cite{DR}.
The basic one-loop tensor integral in $D$ dimensions
for $n$ external particles scattering with outgoing momenta $p_i$,
$n$ internal propagators with masses $M_i$ and $m$ loop momenta
in the numerator can be written,
\begin{displaymath}
\I^D_n[\ell^{\mu_1}\ldots\ell^{\mu_m}]  = \int \frac{d^{D}\ell}{i\pi^{D/2}}
\frac{\ell^{\mu_1}\ldots\ell^{\mu_m}}{
(\ell^2-M_1^2) ((\ell+q_1)^2-M_2^2)
\cdots
((\ell+q_{n-1})^2-M_n^2)},
\end{displaymath}
where $m=1,\ldots,n$ and,
\begin{displaymath}
q_i^\mu = \sum_{j=1}^i p_j^\mu, \qquad\qquad q_0^\mu = q_n^\mu = 0.
\end{displaymath}
The scalar integral is denoted $\I^D_n[1]$.
In the standard approach to such integrals \cite{PV} one utilises the
fact that the tensor structure must be carried by the
external momenta or the metric tensor $g^{\mu\nu}$.
For example, the simplest non-trivial tensor integral
in $D$ dimensions has a single
loop momentum $\ell^{\mu}$.
It reads,
\begin{equation}
\I^D_n[\ell^{\mu}]  \equiv \sum_{j=1}^{n-1} c_j p_j^{\mu},
\label{eq:formfactor}
\end{equation}
using momentum conservation to eliminate one of the momenta.
The formfactors $c_j$ are determined by multiplying both sides by
all possible momenta $p_{i\mu}$ and rewriting $\ell.p_i$ as a difference
of propagator factors,
\begin{displaymath}
\ell.p_i = \frac{1}{2}\left[
((\ell+q_{i})^2-M_{i+1}^2)-((\ell+q_{i-1})^2-M_{i}^2)
+(M_{i+1}^2-M_{i}^2+q_{i-1}^2-q_{i}^2) \right],
\end{displaymath}
thereby
reducing the tensor integral to a sum of scalar integrals,
\begin{displaymath}
\sum_{j=1}^{n-1} c_j p_j.p_i =
\frac{1}{2}\left[
\I^{D,(i+1)}_{n-1}[1]-
\I^{D,(i)}_{n-1}[1]+(M_{i+1}^2-M_{i}^2+q_{i-1}^2-q_{i}^2)\I^{D}_{n}[1]
\right].
\end{displaymath}
Here, $\I^{D,(i)}_{n-1}[1]$ represents the ``pinched''
loop integral with the
$(n-1)$ propagators remaining after
the $i$th propagator factor has cancelled.

The formfactors are then obtained by algebraically solving the
system of $(n-1)$ equations.
This introduces the $(n-1) \times (n-1)$ Gram determinant,
\begin{displaymath}
\Gn = {\rm{det}}(2 p_i \cdot p_j),
\end{displaymath}
(where $i$ and $j$ run over the $(n-1)$ independent momenta),
into the denominator.
Each formfactor is a sum over the scalar integrals present
in the problem multiplied by a kinematic coefficient that
may be singular at the boundary of phase space where the
Gram determinant vanishes.
Typically,
\begin{equation}
c_j \sim \alpha \frac{\I_n^D[1]}{\Delta_n}
+ \sum_{m=1}^n \beta_m \frac{\I_{n-1}^{D,(m)}[1]}{\Delta_n},
\label{eq:toy}
\end{equation}
where the sum is running over the $m$ possible pinchings and
where $\alpha$ and ~$\beta_m$ are coefficients composed
of the kinematic variables.
Since, in many cases, the formfactors $c_j$ are actually finite in
this limit, there are large cancellations
and there may be problems of numerical stability.

The basic approach has been modified in a variety of ways,
including the introduction of
a system of $(n-1)$
reciprocal vectors $v_i^\mu$ (and the associated second rank tensor
$w^{\mu\nu}$ playing the role of $g^{\mu\nu}$)
to carry the tensor structure \cite{OV,EGY,Signer} where,
\begin{displaymath}
v_i^\mu = \epsilon^{p_1\ldots p_{i-1}\mu p_{i+1}\ldots p_n}
\epsilon_{p_1\ldots \ldots p_n}/\Gn,
\end{displaymath}
so that,
\begin{displaymath}
v_i.p_j = \delta_{ij}.
\end{displaymath}
This simplifies the identification of the formfactor coefficients,
but does not eliminate the Gram determinants.
In fact, in both approaches, the number of Gram determinants
generated is equal to the number of loop momenta in the numerator
of the original integral.

A different approach has been suggested by Davydychev \cite{Davy},
who has identified
the formfactors directly as loop integrals in differing numbers
of dimensions and with the loop
propagator factors raised to different powers.
Tarasov \cite{Tar} has obtained recursion relations
for one-loop integrals of this type,
so that a complete reduction is possible.
However, in relating the formfactor loop integrals to
ordinary scalar loop integrals in 4 (or close to 4)
dimensions, the Gram determinant once again appears in the denominator
as in equation~(\ref{eq:toy}).

Finally, Bern, Kosower and Dixon have used the Feynman parameter space
formulation for loop integrals to derive explicit results for the
scalar integrals including the scalar pentagon \cite{BDK1,BDK2}.
The formfactors of the momentum space decomposition are directly
related to Feynman parameter integrals with one or more
Feynman parameters in the numerator.
One can see this by introducing the
auxiliary momentum $\P^\mu$,
\begin{equation}
\P^\mu = -\sum_{i=1}^{n-1} x_{i+1}q_i^\mu,
\label{eq:P}
\end{equation}
so that after integrating out the loop momentum, the tensor
integral for a single loop momentum in the numerator can be expressed
in terms of the external momenta $q_i^\mu$,
\begin{displaymath}
\I^D_n[\ell^{\mu}]  \to
\I^D_n[\P^{\mu}]
\equiv - \sum_{i=1}^{n-1} \I^D_n[x_{i+1}]q_i^\mu.
\end{displaymath}
Here, $\I^D_n[x_i]$ represents the scalar integral with a single factor
of $x_i$ in the numerator.
By comparing with equation~(\ref{eq:formfactor}),
we see that,
\begin{displaymath}
c_j = - \sum_{i=j+1}^{n} \I^D_n[x_i].
\end{displaymath}
Differentiating with respect to the external kinematic variables,
yields relations between integrals with polynomials of
Feynman parameters in the
numerator and the usual scalar integrals.
Once again, the Gram determinant appears in the denominator,
and the final
result for the formfactor $c_j$
combines $n$-point integrals with
the pinched $(n-1)$-point integrals as in equation~(\ref{eq:toy}).

The presence of the Gram determinant is, in some ways, no great surprise.
In the limit $\Gn \to 0$, the $(n-1)$ momenta no longer span an
$(n-1)$-dimensional
space, the $(n-1)$ equations of the Passarino-Veltman approach
are no longer independent and the decomposition is invalid.
Stuart \cite{Stuart} has made modifications
to the basic approach to account for this,
the main observation being that for $\Gn = 0$, the
scalar $n$-point integral can be written as a sum of scalar
$(n-1)$-point integrals.
As a consequence,
there are large cancellations between scalar
integrals with differing numbers
of external legs
in the kinematic limit of vanishing Gram determinants.
For loop corrections to processes
such as quarkonium decay, where the two heavy quarks are
considered to travel collinearly and share the quarkonium momentum,
one can eliminate the Gram determinant singularities
completely using the method of
Stuart \cite{Stuart}.

However, for more general
scattering processes where the collinear limit
may be approached, but is not exact,
the numerical problems as $\Gn \to 0$
remain.
In this paper, we wish to address the problem by combining
the scalar integrals into functions that are well behaved
in the $\Gn \to 0$ limit \cite{GM},
so that the formfactors are given by,
\begin{displaymath}
c_j \sim \alpha \frac{1}{\Delta_n}\left(\I_{n}^{D}[1]+
\sum_{m=1}^n \beta^\prime_m \I_{n-1}^{D,(m)}[1]\right)
+ {\rm finite},
\end{displaymath}
where ``finite'' represents terms that are manifestly
well behaved as
$\Gn \to 0$, and the grouping $(\cdots)$ vanishes with $\Delta_n$.
Such groupings combine
a variety of dilogarithms, logarithms and  constants together
in a non-trivial way.
In fact, for higher rank tensor integrals, with higher powers
of Gram determinants in the denominator,
it becomes even more desirable to organise
the scalar integrals in this way.
It is possible to construct these well behaved groupings by
brute force, making a Taylor expansion of the scalar integrals
in the appropriate limit.   However, as we will show,
they systematically and
naturally arise by considering the scalar integral in
$D+2$ or higher dimensions\footnote{We note that it will not prove
necessary to explicitly compute the scalar integrals in higher dimensions,
since they will be obtained recursively from the
known scalar integrals in $D=4-2\e$ dimensions.}
and/or by differentiating the
scalar integrals
with respect to the external kinematic variables.
Our approach is therefore to re-express the formfactor coefficients
in terms of functions that are finite as $\Gn \to 0$,
explicitly cancelling off factors of the determinant where possible.
The one-loop matrix elements for physical processes will then
depend on these finite combinations, which can themselves
be expanded as a Taylor series to obtain the required numerical precision.
An additional improvement is that the physical size of the resulting
expression is
significantly reduced because the scalar integrals have been combined to
form new, more natural functions.

Of course, one loop amplitudes may also contain spurious singularities
other than those directly arising from Gram determinants.
Such singularities may occur as one or more of the external legs becomes
lightlike or as two external momenta become
collinear.
Our approach has the advantage of avoiding such `fake' singularities.
Since the new finite functions are
obtained by differentiating the scalar integrals, they
cannot contain additional kinematic singularities beyond
those already present in the scalar integral.
This helps to ensure that only genuine poles - those
allowed at tree level - are explicitly present in the one-loop
matrix elements.   Once again, this helps to reduce the size of the
expressions for the amplitudes.

In this paper, we address how such finite functions
might be generated
for arbitrary processes.  We will
closely follow the notation and approach
of Bern, Kosower and Dixon to derive relationships between
integrals with polynomials of Feynman parameters in the numerators
as well as between integrals
with fewer parameters but in higher dimensions.
The basic definitions and notations are introduced in section~2
and the recursive relations for integrals
with up to four Feynman parameters in the numerator are presented
along with the dimension shifting relation of \cite{BDK1,BDK2}.
These expressions are valid for arbitrary internal and external
masses and for general kinematics.
However, making sense of these relations with respect to the
singular limit depends on the actual integral itself; i.e. on
$n$ and
the specific values of the kinematic variables.
The remainder of our paper describes a series of explicit
realisations of the three, four and five point integrals
relevant for the one-loop corrections for the
decay of a virtual gauge boson into four massless partons \cite{GM,DS,BDKW}.
In section~3, we consider
tensor three point integrals for all internal masses equal to zero, but
for general external kinematics.
Section ~4 describes tensor box integrals with
one and two external massive
legs while the tensor pentagon integral is explicitly worked
through in section~5.
Our main results are summarised in section~6, while some explicit
results for three and four point tensor integrals are collected in
the appendices.

\newpage
\section{General notation and basic results}
\setcounter{equation}{0}

The basic integral we wish to work with is the rescaled one-loop integral
in $D$ dimensions\footnote{Note that our
definition of $\I$ differs from that of \cite{BDK1,BDK2} by a factor of
$(-1)^n$.},
\begin{displaymath}
\I^D_n[1] = (-1)^n\Gamma(n-D/2)\int_0^1 d^nx_i \delta(1-\sum_i x_i)
\left[ \sum_{i,j=1}^n S_{ij}x_ix_j \right ]^{D/2-n}.
\end{displaymath}
Here, the Feynman parameters $x_i$ have been introduced and the
loop momentum has been integrated out.
The symmetric matrix $S_{ij}$ contains all the process specific kinematics and
reads,
\begin{displaymath}
S_{ij} = \frac{(M_i^2+M_j^2-(q_{i-1}-q_{j-1})^2)}{2}.
\end{displaymath}
Following closely the steps of \cite{BDK1,BDK2} we perform the projective
transformation \cite{HV},
\begin{displaymath}
x_i = \alpha_i a_i = \frac{\alpha_i u_i}{\sum_{j=1}^n \alpha_j u_j},
\qquad\qquad
\sum_{i=1}^n u_i = 1,
\end{displaymath}
and introduce the constant matrix $\rho_{ij}$ such that,
\begin{displaymath}
S_{ij} = \frac{\rho_{ij}}{\alpha_i\alpha_j}.
\end{displaymath}
The parameters $\alpha_i$ can be related to the kinematic variables present
in the problem, while $\rho_{ij}$ is considered independent of the $\alpha_i$.
Provided that all $\alpha_i$ are real and positive we find,
\begin{equation}
\I^D_n[1] = (-1)^n\Gamma(n-D/2)\int_0^1 d^nu_i \delta(1-\sum_i u_i)
\left (\prod_{j=1}^n \alpha_j \right)
\left (\sum_{j=1}^n \alpha_j u_j \right)^{n-D}
\left[ \sum_{i,j=1}^n \rho_{ij}u_iu_j \right ]^{D/2-n}.
\end{equation}
It is useful to
rescale the integral,
\begin{displaymath}
\I^D_n = \left (\prod_{j=1}^n \alpha_j \right) \Ihat_n^D,
\end{displaymath}
so that in $\Ihat$, the only dependence on the parameters $\alpha_i$
lies in the factor
$\sum_{j=1}^n \alpha_j u_j$.
Differentiating with respect to $\alpha_i$ brings down a factor of the
rescaled Feynman parameter $a_i$ under the integral,
\begin{equation}
\Ihat^D_n[a_i] = \frac{1}{(n-D)}\frac{\partial \Ihat^D_n[1]}{\partial
\alpha_i},
\label{eq:deriv}
\end{equation}
where the notation is obvious.
With repeated differentiation, it is possible to generate all
integrals with Feynman parameters in the numerator.

The second step of Bern, Kosower and Dixon's work \cite{BDK1,BDK2}
is to relate the $n$-point integral with one Feynman parameter in the
numerator to a collection of scalar $n$ and $(n-1)$-point integrals,
\begin{equation}
\Ihat_n^D[a_i] = \frac{1}{2N_n} \sum_{m=1}^n
\left(  \frac{\ghat_i\ghat_m}{\Gnhat} - \eta_{im}
\right)
 \Ihat_{n-1}^{D\,(m)}[1] + \frac{\ghat_i}{\Gnhat}\Ihat_n^D[1],
\label{eq:InxiBDK}
\end{equation}
where\footnote{Note that our definition of $\ghat$ coincides with $\g$ of
\cite{BDK1,BDK2}.},
\begin{displaymath}
\Gnhat = \left(\prod_{i=1}^n \alpha_i \right)^2 \Gn
= \sum_{i,j=1}^n \eta_{ij} \alpha_i \alpha_j
 \equiv \sum_{j=1}^n \alpha_j \ghat_j
\equiv \sum_{j=1}^n \g_j,
\end{displaymath}
and,
\begin{displaymath}
N_n = \frac{1}{2}(\rm{det}~\eta)^{\frac{1}{n-1}}.
\end{displaymath}
In sections~3, 4 and 5, explicit examples using this
notation will be worked through.
Equation~(\ref{eq:InxiBDK}) is the analogue of the
formfactor reduction in momentum space of
\cite{PV} and is easily obtained by integration by parts.
The summation over $m$ represents all possible pinchings of the $n$-point
graph to form $(n-1)$-point integrals.
As expected,
we immediately see the appearance of the Gram determinants in the denominator.
However, equations~(\ref{eq:deriv}) and (\ref{eq:InxiBDK})
are equivalent and since, with a few notable exceptions, the scalar integrals
have a Taylor expansion around $\Gnhat = 0$,
the act of differentiation will not usually
introduce a singular behaviour.
Therefore, we might expect that the $n$-point and
$(n-1)$-point integrals combine
in such a way that the $\Gnhat \to 0$ limit is well behaved.
We can see how this happens by considering the $n$-point integral in
$D+2$ dimensions \cite{BDK1,BDK2,Tar},
\begin{eqnarray}
 \Ihat_n^{D+2}[1] &=&
\frac{1}{(n-1-D/2)(n-D)(n-D-1)}
\rho_{ij}
\frac{\partial^2 \Ihat^D_n[1]}{\partial \alpha_j\partial \alpha_i}
\nonumber \\
&=&
 \ {1\over (n-D-1)} {2N_n\over\Gnhat} \left(\Ihat_n^D[1] \ +\
  {1\over 2N_n} \sum_{m=1}^n \ghat_m \Ihat_{n-1}^{D \, (m)}[1]
    \right),
\label{eq:d-change}
\end{eqnarray}
so that,
\begin{equation}
\Ihat_n^D[a_i]
  \ = \ {1\over 2 N_n} \left( (n-D-1)   \ghat_i   {\Ihat_n^{D+2}}[1]
  \ -  \sum_{m=1}^n \eta_{im} \Ihat_{n-1}^{D\,(m)}[1]\right).
\label{eq:Inai}
\end{equation}
It is important to note that there are no Gram determinants visible
in this equation.   They have all been collected into the higher dimensional
$n$-point integral.
It is clear that if $\Ihat^D_n[1]$ is finite as $\Gnhat \to 0$, then so is
$\Ihat^{D+2}_n[1]$ and
therefore so is $\Ihat_n^D[a_i]$.   This confirms that the apparent divergence
as $\Gnhat \to 0$ is
fake.  Furthermore, $\Ihat^{D+2}_n[1]$ is an excellent candidate for a finite
function - it is well behaved
as the Gram determinant vanishes and is easily related
to the Feynman parameter integrals via equation~(\ref{eq:d-change}).
Of course, it may still be divergent as $\e \to 0$ and the dimensionally
regulated poles remain to be isolated.

By applying the derivative approach, we can easily extend this to two or more
Feynman parameters in the numerator,
\begin{eqnarray}
\Ihat_n^D[a_ia_j]  &=& \frac{1}{(n-D-1)}
\frac{\partial \Ihat^D_n[a_i]}{\partial \alpha_j}  \nonumber \\
&=&
 {1\over 2 N_n} \left(
\ghat_i \frac{\partial {\Ihat_n^{D+2}}[1]}{\partial \alpha_j}
+\frac{\partial \ghat_i}{\partial \alpha_j}   {\Ihat_n^{D+2}}[1]
  \ -  \frac{1}{(n-D-1)}\sum_{m=1}^n \eta_{im}
 \frac{\partial \Ihat_{n-1}^{D\,(m)}[1]}{\partial \alpha_j}\right).
\nonumber
\end{eqnarray}
Using equation~(\ref{eq:deriv}) and the identity,
\begin{displaymath}
\frac{\partial \ghat_i}{\partial\alpha_j} \equiv \eta_{ij},
\end{displaymath}
we see that,
\begin{equation}
\Ihat^D_n[a_i a_j] = \frac{1}{2N_n} \left(
(n-D-2) \ghat_i \Ihat_n^{D+2}[a_j]
+\eta_{ij} \Ihat_n^{D+2}[1]
 - \sum_{m=1}^n \eta_{im}\Ihat^{D\,(m)}_{n-1}[a_j] \right).
\label{eq:Inaiaj}
\end{equation}
Note that $\Ihat^{D\,(m)}_{n-1}[1]$ does not depend on
$\alpha_m$, and therefore,
\begin{displaymath}
\Ihat^{D\,(j)}_{n-1}[a_j] \sim
\frac{\partial \Ihat^{D\,(j)}_{n-1}[1]}{\partial\alpha_j}
\equiv 0.
\end{displaymath}
Consequently, the $m=j$ term in the summation vanishes.

Differentiation
has not produced any new Gram determinants and
we can treat these integrals as new well behaved
building blocks, or
substitute for them using equation~(\ref{eq:Inai})
with $D$ replaced by $D+2$,
\begin{eqnarray}
\Ihat^D_n[a_i a_j] &=& \frac{1}{4N_n^2} \Biggl(
(n-D-2)(n-D-3) \ghat_i\ghat_j \Ihat_n^{D+4}[1]
+2N_n\eta_{ij} \Ihat_n^{D+2}[1]\nonumber \\
&&-(n-D-2)\sum_{m=1}^n \ghat_i\eta_{jm}\Ihat^{D+2\,(m)}_{n-1}[1]
 - 2N_n\sum_{m=1}^n \eta_{im}\Ihat^{D\,(m)}_{n-1}[a_j] \Biggr ).
\nonumber
\end{eqnarray}
The scalar integrals for $D+4$ dimensions can be obtained recursively
from equation~(\ref{eq:d-change}).

Replacing the factors of $\alpha_i$ in equations~(\ref{eq:Inai}) and
(\ref{eq:Inaiaj}) and the analogous equations for three and four
Feynman parameters in the numerator, we find,
\begin{eqnarray}
\I_n^D[x_i]
  &=& {1\over 2 N_n} \left( (n-D-1)   \g_i   {\I_n^{D+2}}[1]
  \ -  \sum_{m=1}^n \eta_{im}\alpha_i\alpha_m \I_{n-1}^{D\,(m)}[1]\right),
\label{eq:Inxi}
\\
\I^D_n[x_i x_j] &=& \frac{1}{2N_n} \Biggl(
(n-D-2) \g_i \I_n^{D+2}[x_j]
+\eta_{ij} \alpha_i\alpha_j\I_n^{D+2}[1]
\nonumber \\
&& \qquad\qquad
 - \sum_{m=1}^n \eta_{im}\alpha_i\alpha_m\I^{D\,(m)}_{n-1}[x_j] \Biggr),
\label{eq:Inxixj}
\\
\I^D_n[x_i x_j x_k ] &=& \frac{1}{2N_n} \Biggl(
(n-D-3) \g_i \I_n^{D+2}[x_j x_k]
+\eta_{ij}\alpha_i\alpha_j \I_n^{D+2}[x_k]
+\eta_{ik}\alpha_i\alpha_k \I_n^{D+2}[x_j]\nonumber \\
&& \qquad\qquad
 - \sum_{m=1}^n \eta_{im}\alpha_i\alpha_m\I^{D\,(m)}_{n-1}[x_jx_k] \Biggr),
\label{eq:Inxixjxk}
\\
\I^D_n[x_i x_j x_k x_l] &=& \frac{1}{2N_n} \Biggl(
(n-D-4) \g_i \I_n^{D+2}[x_j x_k x_l]
+\eta_{ij}\alpha_i\alpha_j \I_n^{D+2}[x_kx_l]
+\eta_{ik}\alpha_i\alpha_k \I_n^{D+2}[x_jx_l]
\nonumber \\
&& \qquad\qquad
+\eta_{il}\alpha_i\alpha_l \I_n^{D+2}[x_jx_k]
 - \sum_{m=1}^n \eta_{im}\alpha_i\alpha_m\I^{D\,(m)}_{n-1}[x_jx_kx_l] \Biggr).
\label{eq:Inxixjxkxl}
\end{eqnarray}
Once again, no Gram determinants are apparent and these
equations may be solved by recursive iteration.
These are our main results
and their use will be made clear
with the explicit examples in the following sections.

Before proceeding to the explicit examples,
we note that the full tensor structure in momentum space is simply
obtained from the Feynman parameter integrals by introducing the
auxiliary momentum $\P^\mu$ defined in equation~(\ref{eq:P}).
With an obvious notation (and after integration of the
loop momentum) the tensor integrals
can be written,
\begin{eqnarray}
\I_n^D[\ell^\mu] &\to& \I_n^D[\P^\mu],\nonumber \\
\I_n^D[\ell^\mu\ell^\nu] &\to& \I_n^D[\P^\mu\P^\nu]-
\frac{1}{2}\I_n^{D+2}[g^{\mu\nu}],\nonumber \\
\I_n^D[\ell^\mu\ell^\nu\ell^\rho] &\to& \I_n^D[\P^\mu\P^\nu\P^\rho]
-\frac{1}{2}\I_n^{D+2}[\{ g\P \}^{\mu\nu\rho}],\nonumber \\
\I_n^D[\ell^\mu\ell^\nu\ell^\rho\ell^\sigma] &\to&
\I_n^D[\P^\mu\P^\nu\P^\rho\P^\sigma]
-\frac{1}{2}\I_n^{D+2}[\{g\P\P\}^{\mu\nu\rho\sigma}]
+\frac{1}{4}\I_n^{D+4}[\{gg\}^{\mu\nu\rho\sigma}],
\nonumber
\end{eqnarray}
where $\{a\ldots b\}^{\mu_1\ldots\mu_n}$ is the usual Passarino-Veltman
notation \cite{PV}, and indicates a sum over all possible
permutations of Lorentz indices carried by $a\ldots b$.
For example,
\begin{displaymath}
\{ g\P \}^{\mu\nu\rho}
= g^{\mu\nu}\P^{\rho}+g^{\nu\rho}\P^{\mu}+g^{\rho\mu}\P^{\nu}.
\end{displaymath}

Throughout the next sections,
we make the simplifying choice that $M_i = 0$.  Such integrals
are relevant for a wide range of QCD processes involving
loops of gluons or massless quarks. The approach can be straightforwardly
extended
to include non-zero masses \cite{BDK2}.

The strategy is to isolate the ultraviolet and infrared
poles from the tensor integrals,
leaving the finite remainder in the form of groups of terms that are well
behaved in all of the kinematic limits. In real calculations where groups of
tensor integrals are combined, this grouping will often cancel as a whole.
Alternatively, if the kinematic coefficient allows, the determinant
can be cancelled off for all of the terms in the function.
This approach is well suited to treatment by an algebraic manipulation program,
once the raw integrals have been massaged to isolate the poles in $\e$
and to group the terms.
As we will show in the explicit examples, this is usually
straightforward.

\section{Three point integrals}
\setcounter{equation}{0}

In processes where the internal lines are massless,
there are only three types of triangle graph
described by the number of massive external legs.
For the one loop corrections to five parton scattering
\cite{BDK5g,BDK2q3g,KST},
only the graphs with one and two massless legs occur.   For processes
involving a gauge boson such as $Z \to 4$~partons \cite{GM,DS,BDKW},
graphs with all external legs massive or off-shell contribute.

\subsection{The three-mass triangle}

\begin{figure}\vspace{6cm}
\includegraphics{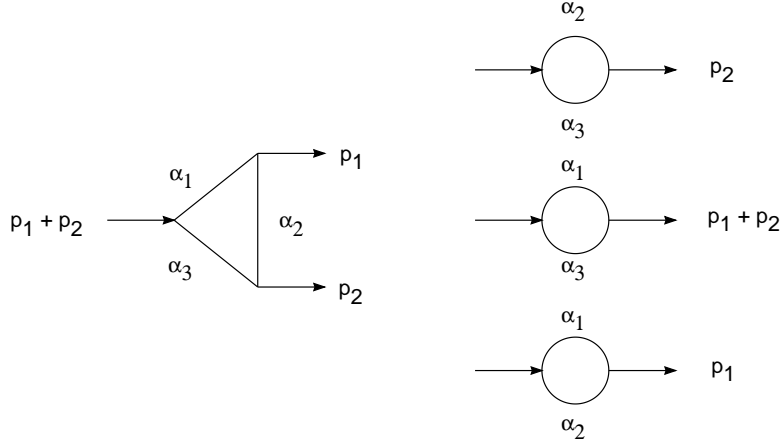}
\caption[]{The triangle graph and each of
the three pinchings obtained by omitting
the internal line associated with $\alpha_m$ for $m=1,2$~and~3.
In each case, the internal line is shrunk to a point and the
momenta at either end are combined.
The relation between the external momenta and the $\alpha_i$ can
be seen by cutting the loop; $\alpha_i\alpha_j = -1/p^2$ where
$p$ is the momentum on one side of the cut and $\alpha_i$,
$\alpha_j$ are associated with the cut lines.}
\label{fig:tripin}
\end{figure}

We first consider triangle integrals with exiting momenta
$p_1$, $p_2$ and $p_3$ as shown in fig.~\ref{fig:tripin}
and all internal masses equal to zero,
$M_i = 0$.
Throughout, we
systematically eliminate $p_3$ (and the Feynman parameter $x_2$)
using momentum conservation so that $p_3 = -(p_1+p_2)$,
$p_3^2 = (p_1+p_2)^2 = s_{12}$ and,
\begin{displaymath}
\P^\mu = -(1-x_1)p_1^\mu - x_3 p_2^\mu.
\end{displaymath}
The full tensor structure with up to three loop momenta in the numerator
can therefore be derived from loop integrals with up to three powers of
$x_1$ or $x_3$ in the numerator.

As a first step, we consider the general case, $p_1^2,~p_2^2,~s_{12}  \neq 0$,
where the scalar integral in four dimensions is known to be finite.
Here the $\alpha_i$ parameters can be determined by,
\begin{displaymath}
\alpha_1\alpha_2   p_1^2=-1, \qquad \alpha_2\alpha_3
p_2^2=-1, \qquad \alpha_1\alpha_3 s_{12} =-1,
\end{displaymath}
while,
\begin{eqnarray}
\Delta_3  &=& -p_{1}^4-p_{2}^4-s_{12}^2+2p_1^2p_2^2
		   +2p_1^2s_{12}+2p_2^2s_{12},\nonumber \\
\G3hat &=& -\alpha_1^2-\alpha_2^2-\alpha_3^2 + 2\alpha_1\alpha_2
+ 2\alpha_1\alpha_3+ 2\alpha_2\alpha_3.
\nonumber
\end{eqnarray}
{}From the definition of the matrix $\eta$, we see that,
\begin{displaymath}
\eta_{ij} =
\left(
\begin{array}{rrrr}
-1 & 1 & 1 \\ 1 & -1 & 1 \\ 1 & 1 & -1
\end{array}
\right),
\qquad N_3 = 1.
\end{displaymath}
The variables $\g$ always appear in the following combinations,
\begin{displaymath}
\frac{\gamma_i}{\prod_{j=1}^3 \alpha_j^2} =
\left\{
\begin{array}{ll}
p_2^2(p_1^2+s_{12}-p_2^2) & i=1 \\
s_{12}(p_2^2+p_1^2-s_{12}) & i=2 \\
p_1^2(p_2^2+s_{12}-p_1^2) & i=3 \\
\end{array}
\right. .
\end{displaymath}

The scalar triangle integral for all external masses non-zero
is finite in four dimensions \cite{3pt,BDK1,EGY}
,\footnote{For scalar integrals in
$D=4-2\epsilon$ or $D=4$ dimensions, we omit the superscript $D$.}
\begin{equation}
\I_3[1]
=\frac{1}{\sqrt{-\Delta_3}}
\left(\log(a^+a^-)\log\left(\frac{1-a^+}{1-a^-}\right)
+2\Li2(a^+)-2\Li2(a^-)\right),
\label{eq:I3}
\end{equation}
where $\Li2$ is the usual dilogarithm function and
$a^\pm $ are two roots of a quadratic equation,
\begin{displaymath}
a^{\pm}=\frac{s_{12}+p_2^2-p_1^2 \pm \sqrt{-\Delta_3}}{2 s_{12}}.
\end{displaymath}
Although $\I_3[1]$ appears to diverge as $\Delta_3 \to 0$, this is not the
case.
As noted by Stuart \cite{Stuart}, in this limit, the triangle graph reduces
to a sum of bubble graphs,
\begin{displaymath}
\lim_{\Delta_3 \rightarrow 0} \I_3[1] =
\frac{2}{s_{12}+p_1^2-p_2^2}\log\left(\frac{s_{12}}{p_2^2}\right)
+\frac{2}{s_{12}+p_2^2-p_1^2}\log\left(\frac{s_{12}}{p_1^2}\right),
\end{displaymath}
and there is a well behaved Taylor series in $\Delta_3$.

\subsubsection{Tensor integrals in $D=4$}

The tensor integral can be easily written in terms of higher dimensional
scalar integrals and bubble scalar integrals using
eqs.~(\ref{eq:Inxi}--\ref{eq:Inxixjxk}).
For one Feynman parameter in the numerator, this gives,
$$
\I_3^{D=4-2\e}[x_i] =   -(1-\e) \g_i \I_3^{D=6-2\e}[1] - \frac{1}{2}
\sum_{m=1}^3 \eta_{im} \alpha_i \alpha_m \I_2^{D=4-2\e (m)}[1]  .
$$
Immediately a problem is apparent -- the coefficient of the
scalar integral in $( 6-2\e ) $ dimensions, $\gamma_i$, is singular as one or
more of the external momenta become lightlike.
Although the divergence as the Gram
determinant vanishes has been removed, it appears to have been replaced by a
divergence
as the invariants vanish\footnote{Problems in this limit are to be
expected since even the scalar integral itself is not finite as
$p_i^2 \to 0$.}. However, these divergences cancel between the triangle and
bubble contributions and the tensor integral itself is well behaved and finite
in all kinematic limits and is therefore a
better choice for a finite function.

In fact, since the triangle scalar integral is finite in $4$ dimensions,
it is convenient to generate the tensor structure directly from
derivatives of the scalar integral.
However, in order
to use equation~(\ref{eq:InxiBDK}), we also need the
two point integral for external momentum $p$ (and internal masses $M_i = 0$),
\begin{equation}
\I^D_2[1] = \frac{\Gamma(2-D/2)\Gamma^2(D/2-1)}{\Gamma(D-2)} (-p^2)^{D/2-2},
\label{eq:I2}
\end{equation}
for each of the three pinchings, $m = 1,2$ and 3 shown in
fig.~\ref{fig:tripin}.
For pinching $m$ and $D=4-2\e$,
\begin{displaymath}
\Ihat_2^{(m)}[1] = \left(\frac{\alpha_m}{\prod_{i=1}^3 \alpha_i}\right)
\I_2^{(m)}[1]
=
\frac{c_\Gamma}{\e (1-2\e)}
\left(\frac{\alpha_m}{\prod_{i=1}^3 \alpha_i}\right)^{1-\e},
\end{displaymath}
where the usual product of Gamma functions obtained in one-loop integrals
$c_\Gamma$ is given by,
\begin{displaymath}
c_\Gamma = \frac{\Gamma^2(1-\e)\Gamma(1+\e)}{\Gamma(1-2\e)}.
\end{displaymath}
Rewriting equation~(\ref{eq:InxiBDK}) for the case $D=4$, $n=3$
and $i=3$
and adding,
\begin{displaymath}
\frac{1}{2}
\sum_{m=1}^3 \left( \eta_{3m} - \frac{\ghat_3\ghat_m}{\G3hat} \right)
\frac{\alpha_m}{\alpha_2}\Ihat_2^{(2)}[1]
 = 0,
\end{displaymath}
we see,
\begin{eqnarray}
\Ihat_3[a_3]
&=&
\frac{1}{\G3hat }
\left (  \ghat_3 \Ihat_3[1]
+\frac{1}{2}
 \sum_{m=1}^3
\left(  \ghat_3\ghat_m - \eta_{3m}\G3hat
\right)
\frac{\alpha_m}{\prod \alpha}
\left ( \I_{2}^{(m)}[1] - \I_{2}^{(2)}[1] \right) \right)\nonumber \\
&=&
\frac{1}{\G3hat }
\left (  \ghat_3 \Ihat_3[1]
-\frac{\ghat_2}{\alpha_3}\log\left(\frac{\alpha_2}{\alpha_1}\right)
+2\log\left(\frac{\alpha_2}{\alpha_3}\right) \right) + {\cal O}(\e).
\end{eqnarray}
Alternatively, this could be obtained by differentiation of
equation~(\ref{eq:I3}).
By trivial replacement of factors of $\alpha$,
we find,
\begin{eqnarray}
\I_3[x_3]
&=&
\frac{1}{\Delta_3}
\Biggl(p_1^2(s_{12}+p_2^2-p_1^2)\I_3[1]
    +(p_1^2+p_2^2-s_{12})\log\left(\frac{s_{12}}{p_2^2}\right)
            -2p_1^2\log\left(\frac{s_{12}}{p_1^2}\right)\Biggr ).
\label{eq:I3x3}
\end{eqnarray}
Integrals with higher powers of Feynman parameters can now be generated
by direct differentiation of $\Ihat_3[a_3]$,
\begin{displaymath}
\I_3[x_ix_3^n] = -\frac{1}{(n+1)} \left(\prod \alpha \right )
\alpha_i\alpha_3^n
\frac{\partial}{\partial \alpha_i} \Ihat_3[a_3^n].
\end{displaymath}
All of these functions will be finite in the $\e \to 0$ limit, and can
be considered as building blocks in constructing the tensor structures for
box and pentagon integrals.
In fact, because they are obtained by differentiating
a function well behaved as $\Delta_3 \to 0$,
they are also finite in this limit.
Therefore, they tie together the dilogarithms from the
triangle integrals and the logarithms from bubble integrals in an
economical and numerically very stable way.

We also see that they are directly generated in tensor structures
for box graphs (equations~(\ref{eq:Inxi}--\ref{eq:Inxixjxkxl}) with $n=4$)
and will naturally cancel in Feynman diagram calculations involving
both triangle and box graphs.
For general calculations with $p_1^2 \neq 0$ and $p_2^2 \neq 0$,
we introduce the notation,
\begin{equation}
\Lc{0}(p_1,p_2) = \I_3[1],
\qquad
\Lc{2n-1}(p_1,p_2) = \I_3[x_3^n],
\qquad
\Lc{2n}(p_1,p_2) = \I_3[x_1x_3^n],
\end{equation}
for $\Lc{0\ldots 5}$.
The symmetry properties of the triangle function imply that the analogous
functions for $x_1 \leftrightarrow x_3$ (or $\alpha_1 \leftrightarrow
\alpha_3$)
are just obtained by exchanging $p_1$ and $p_2$.
In dealing with box graphs, integrals with $x_2$ in the numerator will
naturally
arise.  In these cases, we systematically eliminate them using $\sum_i x_i =
1$.
Explicitly, we find,
\begin{eqnarray}
\I_3[x_1x_3] &=&
\frac{1}{2\Delta_3}       \Biggl(
2p_2^2(s_{12}+p_1^2-p_2^2)\I_3[x_3]
+p_1^2(s_{12}+p_2^2-p_1^2)\I_3[x_1] \nonumber \\
&&\qquad\qquad
 -p_1^2p_2^2 \I_3[1]
        - p_2^2\log\left(\frac{s_{12}}{p_2^2}\right)  +p_1^2+p_2^2-s_{12}
\Biggr ),
\label{eq:I3x1x3}
\\
\I_3[x_3^2] &=&
\frac{1}{2\Delta_3}       \Biggl(
3p_1^2(s_{12}+p_2^2-p_1^2)\I_3[x_3]
 +p_1^4 \I_3[1]
        - (s_{12} -p_2^2)\log\left(\frac{s_{12}}{p_2^2}\right)  -2p_1^2 \Biggr
),
\label{eq:I3x3x3}
\\
\I_3[x_1x_3^2] &=&
\frac{1}{6\Delta_3}       \Biggl(
4p_2^2(s_{12}+p_1^2-p_2^2)\I_3[x_3^2]
+6p_1^2(s_{12}+p_2^2-p_1^2)\I_3[x_1x_3] \nonumber \\
&&\qquad\qquad
-3p_1^2p_2^2 \I_3[x_3]
+p_1^4 \I_3[x_1]
         -p_2^2\log\left(\frac{s_{12}}{p_2^2}\right)
+p_2^2-s_{12} \Biggr ),
\label{eq:I3x1x3x3}
\\
\I_3[x_3^3] &=&
\frac{1}{3\Delta_3}       \Biggl(
5p_1^2(s_{12}+p_2^2-p_1^2)\I_3[x_3^2]
 +2p_1^4 \I_3[x_3]
        - (s_{12} -p_2^2)\log\left(\frac{s_{12}}{p_2^2}\right)  - p_1^2 \Biggr
).
\label{eq:I3x3x3x3}
\end{eqnarray}

\begin{figure}\vspace{8cm}
\includegraphics{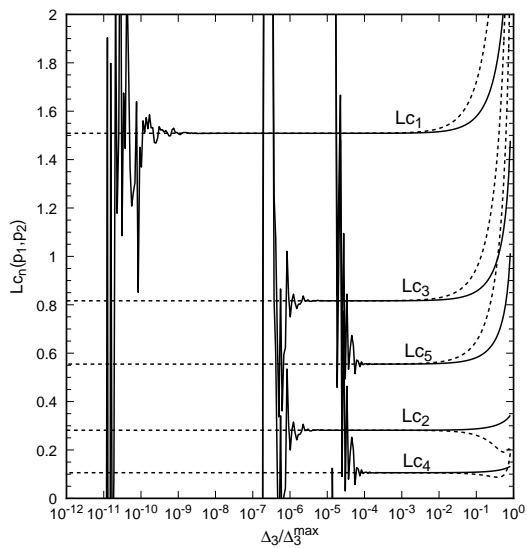}
\caption[]{The finite functions for the triply massive triangle graph
with $s_{12} = 1,~~p_1^2 = 0.2$ as a function of
$\Delta_3/\Delta_3^{{\rm max}}$ where $\Delta_3^{{\rm max}} =
-(s_{12}-p_1^2)^2$.
The functions have been evaluated using double precision Fortran.
The dashed lines show the
approximate form for the function in the limit $\Delta_3 \to 0$,
retaining only the first term of the
Taylor expansion.}
\label{fig:lcn}
\end{figure}

By expanding as a series in $\Delta_3$, these functions
can be evaluated
near the singularity with arbitrary precision.
For example,
\begin{eqnarray}
\lim_{\Delta_3 \rightarrow 0} \I_3[x_3^2] &=&
\frac{(s_{12}+p_1^2-p_2^2)}{3s_{12}(s_{12}-p_1^2-p_2^2)}
-\frac{2p_1^4}{3s_{12}(s_{12}-p_1^2-p_2^2)^2}
\log\left(\frac{s_{12}}{p_1^2}\right)\nonumber \\
&&\qquad
+\frac{(3s_{12}-3p_1^2+p_2^2)}{6s_{12}(s_{12}-p_1^2-p_2^2)}
\log\left(\frac{s_{12}}{p_2^2}\right).\nonumber
\end{eqnarray}
To illustrate this, fig.~\ref{fig:lcn} shows the various functions at a
specific phase space point, \linebreak
$s_{12} = 1,~~p_1^2 = 0.2$ and letting $p_2^2$ vary in such
a way that $\Delta_3 \to 0$.
This corresponds to $p_2^2 \to 0.135$.
We  see that as this limit is approached, the numerical evaluation
of the function using double precision (an
intrinsic numerical precision, $acc$,  of roughly $10^{-14}$)
becomes uncertain.
For this particular phase space point, functions with a single
Gram determinant in the denominator ($\Lc{1}$) remain stable until
$\Delta_3/\Delta_3^{{\rm max}} \sim 10^{-9}$ while those with more powers
of the Gram determinant break down correspondingly sooner - at
$\Delta_3/\Delta_3^{{\rm max}} \sim 10^{-6}$ for $\Lc{2}$ and $\Lc{3}$ and
$\Delta_3/\Delta_3^{{\rm max}} \sim 10^{-4}$ for $\Lc{4}$ and $\Lc{5}$.
In general, numerical problems typically occur
when $\Delta_3/\Delta_3^{{\rm max}} \sim (acc)^{1/N}$ where $N$ is the
number of Gram determinants in the denominator of the function.
Other phase space points yield a similar behaviour.

The unstable points represent a rather small proportion of the
allowed phase space.
However, problems may arise
using adaptive Monte Carlo methods such as VEGAS \cite{VEGAS} where the phase
space is preferentially sampled where the matrix elements are large.
Finding an anomalously high value for the matrix elements in a region of
instability would cause the Monte Carlo integration to focus on that region
giving unpredictable results.

Of course, these instabilities could be handled by a brute force
increase in numerical precision.  While possible, this has the disadvantage
of producing significantly slower code, and, since in all cases, the
approximate form obtained by making a Taylor expansion
about $\Delta_3 = 0$ and keeping only the constant term  works well where the
numerical instabilities begin, this is not an attractive solution.
In fact, the approximation is reliable for $\Delta_3 < 10^{-3}
 \Delta_3^{{\rm max}}$.
Explicit forms for the approximations are collected in Appendix~C.

\subsubsection{Scalar integrals in higher dimensions}

We now turn to the scalar triangle integrals in higher dimensions.
They appear in the $g_{\mu\nu}$
part of the general Lorentz structure and recursively in
the determination of $\I_4^{D=4-2\e}[x_i...x_l]$.
Unlike the triangle in four dimensions, these integrals are
ultraviolet divergent due to the presence of
the various pinchings - bubble integrals.
A function that can usefully be used as a building block of matrix element
calculations, must be finite as both $\Delta_3 \to 0$ {\em and} $\e \to 0$
and we must first isolate the poles in $\e$.
Although equation~(\ref{eq:d-change}) suggests that
the ultraviolet pole structure involves $\Delta_3$,
this is easily shown not to be the case.
Adding  terms proportional to,
\begin{displaymath}
\frac{1}{2}\left( \G3hat - \sum_{m=1}^3 \ghat_m \alpha_m\right)
\frac{1}{\alpha_2} \Ihat^{(2)}_2[1] = 0,
\end{displaymath}
to equation~(\ref{eq:d-change}) for $n=3$, $D=4-2\e$ we see,
\begin{displaymath}
\Ihat^{D=6-2\e}_3[1] =
-\frac{1}{(1-\e)}\frac{1}{\G3hat}
\left (\Ihat_3[1] +
\frac{1}{2}
\sum_{m=1}^3 \frac{\g_m}{\prod \alpha}
\left(\I_2^{(m)}[1] - \I_2^{(2)}[1]\right)
+ \frac{\G3hat}{2\prod \alpha} \I_2^{(2)}[1]\right),
\end{displaymath}
where the divergence as $\e \to 0$ lies exclusively in the last term.
Reinserting the factors of $\alpha$ and the definition of
$\I_2[1]$ in $D=4-2\e$ dimensions we find,
\begin{equation}
\I^{D=6-2\e}_3[1] =
\Lc{1S}(p_1,p_2)
-\frac{1}{2}\left(\frac{(-s_{12})^{-\e}}{\e}+3\right) c_\Gamma,
\label{eq:I36}
\end{equation}
where,
\begin{eqnarray}
\Lc{1S}(p_1,p_2) &=& \frac{1}{2\Delta_3}
\Biggl (2p_1^2p_2^2s_{12}\I_3[1]
-p_1^2(s_{12}+p_2^2-p_1^2)
\log\left(\frac{s_{12}}{p_1^2}\right)\nonumber \\
&&\qquad -p_2^2(s_{12}+p_1^2-p_2^2)
\log\left(\frac{s_{12}}{p_2^2}\right)\Biggr)\nonumber \\
&=& \frac{1}{2} \left (p_1^2 \I_3[x_1] + p_2^2 \I_3[x_3] \right).
\label{eq:Lc1s}
\end{eqnarray}

In a similar fashion, the $\e$ pole structure can be removed from the
triangle scalar integral in $D=8-2\e$ and $D=10-2\e$
dimensions yielding two more functions that are finite as both
$\Delta_3\to 0$ and $\e \to 0$.  Explicitly we find,
\begin{eqnarray}
\I^{D=8-2\e}_3[1] &=&
\Lc{2S}(p_1,p_2) -\frac{(p_1^2+p_2^2+s_{12})}{24}
\left(\frac{(-s_{12})^{-\e}}{\e} + \frac{19}{6}\right) c_\Gamma,
\label{eq:I38}
\\
\I^{D=10-2\e}_3[1] &=&
\Lc{3S}(p_1,p_2)
-\frac{(p_1^4+p_2^4+s_{12}^2+p_1^2p_2^2+p_1^2s_{12}+p_2^2s_{12})}{360}
\left(\frac{(-s_{12})^{-\e}}{\e}+\frac{17}{5}\right) c_\Gamma,\nonumber \\
\label{eq:I310}
\end{eqnarray}
where the finite functions are defined by,
\begin{eqnarray}
\Lc{2S}(p_1,p_2) &=&
\frac{1}{4\Delta_3} \left( 2p_1^2p_2^2s_{12} \Lc{1S}(p_1,p_2)
	   -\frac{1}{6} \left( p_1^4(s_{12}+p_2^2-p_1^2)
	   \log\left(\frac{s_{12}}{p_1^2}\right) \right. \right. \nonumber \\
&&\qquad  \left. \left. + p_2^4(s_{12}+p_1^2-p_2^2)
	   \log\left(\frac{s_{12}}{p_2^2}\right)
	   +2p_1^2p_2^2s_{12} \right) \right),
\label{eq:Lc2s}
\\
\Lc{3S}(p_1,p_2) &=&
\frac{1}{6\Delta_3}
\Biggl (2p_1^2p_2^2s_{12}\Lc{2S}(p_1,p_2)
-\frac{1}{60}\Biggl (p_1^6(s_{12}+p_2^2-p_1^2)
\log\left(\frac{s_{12}}{p_1^2}\right)\nonumber \\
&&\qquad
+p_2^6(s_{12}+p_1^2-p_2^2)
\log\left(\frac{s_{12}}{p_2^2}\right)
+\frac{p_1^2p_2^2s_{12}}{2}(p_1^2+p_2^2+s_{12})\Biggr)\Biggr).
\label{eq:Lc3s}
\end{eqnarray}

Although these functions have been obtained via equation~(\ref{eq:d-change}),
they are still related to derivatives of the basic scalar integral in
$D=4-2\e$,
and are therefore finite in the $\Delta_3 \to 0$ limit.
We can see this by examining the
same phase space point as before, $s_{12} = 1,~~p_1^2 = 0.2$
and varying $\Delta_3$.
As expected,
the $\Lc{nS}$ show a similar behaviour to the $\Lc{n}$ functions
- numerically breaking down at larger and larger values of $\Delta_3$
as the number of Gram determinants increases, and
being well described by the first term in the Taylor expansion
as this happens.
For completeness, we collect the limiting approximations
in Appendix~C.

\subsubsection{Tensor integrals in higher dimensions}

For triangle loop integrals with three loop momenta in the numerator, it is
also necessary to know the $D=6-2\e$ integral with a single Feynman parameter
in the numerator.
Rather than differentiating the ultraviolet divergent $\I_3^{D=6-2\e}[1]$,
we can evaluate it in terms of the $D=4-2\e$ tensor integrals of section~3.1.1.
Using (\ref{eq:Inaiaj}) for $D=4-2\e$, we see that,
\begin{eqnarray}
\Ihat_3[a_1a_j]+\Ihat_3[a_3a_j]
&=& \frac{1}{2}\Biggl(
-(3-2\e)(\ghat_1+\ghat_3)\Ihat_3^{D=6-2\e}[a_j]
+(\eta_{1j}+\eta_{3j})\Ihat_3^{D=6-2\e}[1]\nonumber \\
&&\qquad
-\sum_{m=1}^3 (\eta_{1m}+\eta_{3m})\Ihat_2^{(m)}[a_j] \Biggr),
\nonumber
\end{eqnarray}
which, for $j=1,3,$\footnote{Since the
Feynman parameters add to one, the case
$j=2$ is of little interest.} simplifies using,
\begin{displaymath}
\eta_{1j}+\eta_{3j} = 0.
\end{displaymath}
The same equation simplifies the sum over bubble pinchings so that only $m=2$
contributes,
while $\ghat_1 + \ghat_3 = 2\alpha_2$.
Restoring the factors of $\alpha$ and using,
\begin{displaymath}
\I_2^{(2)}[x_1] = \I_2^{(2)}[x_3] = \frac{1}{2} \left( \frac{(-s_{12})^{-\e}}
 {\e} + 2 \right)c_\Gamma,
\end{displaymath}
yields,
\begin{equation}
\I_3^{6-2\e}[x_j] = \frac{1}{3} \Biggl( p_1^2 \I_3[x_1 x_j] + p_2^2
 \I_3[x_3 x_j] \Biggr) - \frac{1}{6} \left( \frac{(-s_{12})^{-\e}} {\e} +
\frac{8}{3}
\right)c_\Gamma .
\end{equation}

Later, we will see that constructing the tensor integrals for box graphs can
also
generate $\I_3^{6-2\e}[x_j]$
 and higher dimensional triangle integrals with more parameters in the
numerator.
In each case, we use similar tricks with
equations~(\ref{eq:Inxi}--\ref{eq:Inxixjxkxl}) to
rewrite them in terms of the four dimensional
integrals with the ultraviolet pole made explicit.

\subsection{The two-mass triangle}

We will also be interested in triangle graphs where
one or more of the external momenta is lightlike.
Here, we first focus on the case, $p_2^2 \to 0$.
In $D=4-2\e$ dimensions, we have the well known result,
\begin{equation}
\Ihat_3^{2m}[1] = \frac{c_\Gamma}{\e^2}  \left(
\frac{ (\alpha_1\alpha_3)^\e -  (\alpha_1\alpha_2)^\e}{\alpha_3-\alpha_2}
\right ),
\end{equation}
where the superscript indicates that only two of the three legs are massive.
For this choice of kinematics,
\begin{displaymath}
\alpha_1\alpha_2 p_1^2 = -1,\qquad \alpha_1\alpha_3 s_{12} = -1,
\end{displaymath}
and
\begin{displaymath}
\G3hat = -(\alpha_3-\alpha_2)^2,
\end{displaymath}
so that the singular limit is $\alpha_3 \to \alpha_2$.
Because $\G3hat$ makes no reference to $\alpha_1$,
$\eta_{ij}$ contains a row of zeroes,
\begin{displaymath}
\eta_{ij} =
\left(
\begin{array}{rrrr}
0 & 0 & 0 \\ 0 & -1 & 1 \\ 0 & 1 & -1
\end{array}
\right),
\end{displaymath}
and therefore $N_3 = 0$.
Consequently, care is needed in
applying the equations of section~2.
In addition, since the scalar integral for three massive legs is finite (and
the
results in the preceding subsections have been explicitly derived in $D=4$),
one cannot just set $p_2^2 \to 0$.

In fact, it is easiest just to bypass the problem and
generate the whole tensor structure
by direct differentiation of the scalar integral
with respect to $\alpha_3$ and $\alpha_1$.
It is easy to see that,
\begin{displaymath}
\Ihat_3^{2m}[a_3^n] \sim
A \Ihat_3^{2m}[1] + B \frac{(\alpha_1\alpha_3)^\e}{\e}
+ C \log\left(\frac{\alpha_3}{\alpha_2}\right)
+ D,
\end{displaymath}
where $A,B,C$ and $D$ are polynomials in $1/\alpha_3$ and
$1/(\alpha_3-\alpha_2)$.
Unlike the all massive case discussed before, the scalar integral is
singular as $\e \to 0$.
As a general rule it is not necessary to be
particularly careful with double poles in $\e$, since
they must either cancel or form the infrared poles of
real matrix elements.   However, it is possible for the integrals to be
multiplied by factors of $\e$ - from expanding factors of dimension -
and the resulting logarithms
should occur in combinations that are finite as $\alpha_3 \to \alpha_2$.
It is easy to see that,
\begin{displaymath}
\e \times  \Ihat_3^{2m}[1] = \frac{\log\left(\frac{\alpha_3}{\alpha_2}\right)}
{\alpha_3-\alpha_2} + {\cal O}(\e),
\end{displaymath}
is finite.
So, to tie the logarithms and constants together in combinations that
are well behaved in the $\alpha_3 \to \alpha_2$ limit,
we use the fact that derivatives of this function are also well behaved,
and introduce the functions,
\begin{equation}
\Lc{n}^{2m}(p_1,p_2) = -  \lim_{\e \rightarrow 0} \left( \e
			\times \I^{2m}_3[x_3^{n-1}] \right),
\end{equation}
for $n=1,\ldots,4$.
In terms of invariants,
\begin{equation}
\Lc{n}^{2m}(p_1,p_2) = -\left(\frac{p_1^2\Lc{n-1}^{2m}(p_1,p_2) -
\frac{1}{n-1}}{s_{12}-p_1^2}\right), \qquad n \geq 2
\label{eq:lcmassless}
\end{equation}
with,
\begin{equation}
\Lc{1}^{2m}(p_1,p_2)=
\frac{\log\left(\frac{s_{12}}{p_1^2}\right)}{s_{12}-p_1^2},
\qquad
\Lc{0}^{2m}(p_1,p_2)=
\log\left(\frac{s_{12}}{p_1^2}\right).
\label{eq:lcmassless1}
\end{equation}
These functions, or functions closely related to them,
have appeared in
next-to-leading order matrix element
calculations~\cite{BDK5g,BDK2q3g,KST,Signer}.
The explicit forms for $\I_3^{2m}[x_ix_j]$ appearing in the
momentum expansion are well known and are collected in Appendix A.

\begin{figure}\vspace{8cm}
\includegraphics{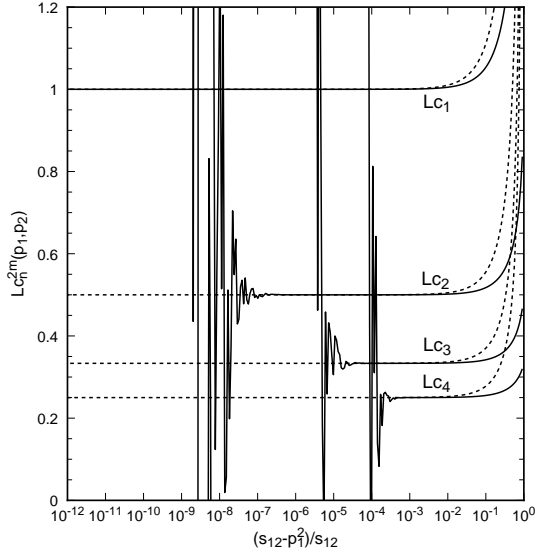}
\caption[]{The finite functions for the triangle graph with two external masses
with $s_{12} = 1$ evaluated in double precision Fortran as a function of
$(s_{12}-p_1^2)/s_{12}$.
The dashed lines show the
approximate form for the function in the limit $p_1^2 \to s_{12}$,
retaining only the first term of the
Taylor expansion.}
\label{fig:lcn00}
\end{figure}

Although these functions are rather simple, they still contain numerical
instabilities
as $p_1^2 \to s_{12}$.  This can be seen in Fig.~\ref{fig:lcn00} where we show
$\Lc{n}^{2m}$ for the specific phase space point $s_{12} = 1$ and let $p_1^2$
approach
$s_{12}$.   While a single inverse powers of
$(s_{12}-p_1^2)$ is handled correctly,
higher powers cause problems.
As can be seen from the figure, a suitable approximation is obtained by the
first
term in the Taylor expansion,
\begin{displaymath}
\lim_{p_1^2 \to s_{12}} \Lc{n}^{2m}(p_1,p_2)  = \frac{1}{n p_1^2},
\end{displaymath}
for $n \geq 1$.

The other configuration of triangle graph that appears is where two of the
momenta are lightlike, $p_1^2 = p_2^2 = 0$.
Once again, the tensor structure can be generated by differentiation or
canonical Passarino-Veltman reduction.  Here, there is only
one scale in the problem so there can be no logarithms and it is neither
possible
nor necessary to introduce well behaved functions.
The explicit forms for the Feynman parameter integrals appearing in the
momentum expansion are well known and for the sake of completeness
are given in Appendix A.

This concludes our discussion of triangle graphs.  For the case of three
massive external legs (and internal masses set equal to zero)
the four dimensional tensor
integrals are finite as $\e \to 0$ and
are given by the functions $\Lc{0\ldots 5}(p_1,p_2)$ defined in
(\ref{eq:I3},\ref{eq:I3x3},\ref{eq:I3x1x3}--\ref{eq:I3x3x3x3}),
while the ultraviolet divergent part
$g_{\mu\nu}$ part, $\I_3^{D=6-2\e}[1]$ is expressed
in terms of a similar function ($\Lc{1S}$) with
the pole isolated (\ref{eq:Lc1s}).
For the tensor structure in the simpler case with one lightlike leg,
it is useful to group the logarithms and constants
using the functions $\Lc{1\ldots 4}^{2m}(p_1,p_2)$ (\ref{eq:lcmassless1},
\ref{eq:lcmassless}).

For the more general case where the internal masses are non-zero, the
same procedure can be utilised. The matrix
$\eta$ has slightly more entries and there are more scales in the problem.
However,
the grouping together of triangle graphs and bubble integrals
into functions
well behaved in the $\Delta_3 \to 0$ limit and the isolation of
the ultraviolet singularities
can be made explicit in the same way.

\section{Four point integrals}
\setcounter{equation}{0}

\begin{figure}\vspace{8cm}
\includegraphics{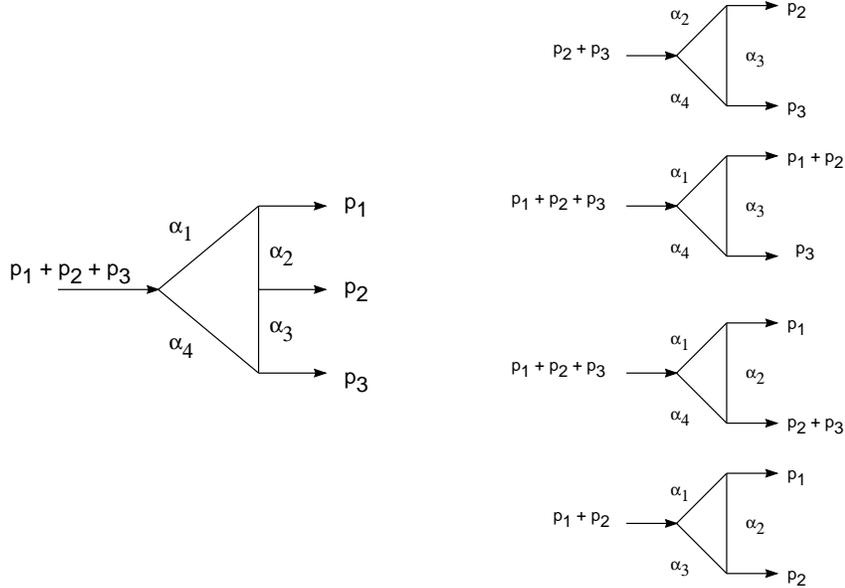}
\caption[]{The box graph and each of
the four pinchings obtained by omitting
the internal line associated with $\alpha_m$ for $m=1,2,3$~and~4.}
\label{fig:boxin}
\end{figure}

For one-loop corrections to five parton scattering,
box graphs with at most one external leg occur.
However, for processes involving a massive vector boson and
four massless partons, we can obtain box graphs with a second massive
external leg by pinching together two of the partons.
There are two distinct configurations according to the positions
of the massive legs; the adjacent box graph and the opposite box graph.
The box graph is shown in fig.~\ref{fig:boxin} for
outgoing momenta $p_1$, $p_2$ and $p_3$.
Throughout this section, we will assume that $(p_1+p_2+p_3)^2 = s_{123} \neq
0$.
In the adjacent two mass case, $p_2^2 = p_3^2 = 0$ and $p_1^2 \neq 0$,
while for the opposite box,  $p_1^2 = p_3^2 = 0$ and $p_2^2 \neq 0$.
Unfortunately, the raw scalar integrals for these two cases behave rather
differently.
The adjacent box is finite in the limit that $\Delta_4\to 0$,
while the opposite box diverges as
$\Delta_4\to 0$.
In this section, we work through these two configurations and rewrite the
tensor
integrals in terms of well behaved functions and explicit poles in $\e$.

\subsection{The adjacent two-mass box}

We first consider the adjacent box with $p_2^2 = p_3^2 = 0$
and all internal masses equal to zero.
As in the triangle case, we systematically eliminate one of the momenta,
$p_4 = -(p_1+p_2+p_3)$, and one of the Feynman parameters,
$x_3 = 1-x_1-x_2-x_4$,
so that,
\begin{displaymath}
\P^\mu = -(1-x_1)p_{12}^\mu + x_2 p_2^\mu - x_4 p_3^\mu.
\end{displaymath}
The related integrals with $p_3^2 \neq 0$ and $p_1^2 = 0$
are obtained by $p_1 \leftrightarrow p_3$ (and the indices $i$ and $j$ in
$\alpha_i$ and $\eta_{ij}$ transform as $1 \leftrightarrow 4$ and
$2 \leftrightarrow 3$).

For this kinematic configuration, the $\alpha$ parameters are defined by,
\begin{displaymath}
\begin{array}{rr}
\alpha_1\alpha_4 \, s_{123}=-1, & \alpha_1\alpha_2 \, p_1^2 =-1 \\
  \alpha_1\alpha_3 \, s_{12}=-1, & \alpha_2\alpha_4 \, s_{23}=-1,
\end{array}
\end{displaymath}
while,
\begin{displaymath}
\begin{array}{lr}
\eta_{ij} =
\left(
\begin{array}{rrrr}
0 & 0 & 1 & 0 \\ 0 & 0 & -1 & 1 \\
1 & -1 & 2 & -1 \\ 0 & 1 & -1 & 0
\end{array}
\right),
& N_4 = \frac{1}{2},
\end{array}
\end{displaymath}
and,
\begin{displaymath}
\Delta_4 = 2s_{23}\left( (s_{123}-s_{12})(s_{12}-p_1^2)-s_{12}s_{23}\right).
\end{displaymath}
The coefficients $\g_i$ always appear in the following combinations which are
directly related to the conventional variables,
\begin{displaymath}
\frac{\gamma_i}{  \prod_{j=1}^4 \alpha_j} =
\left\{
\begin{array}{ll}
-s_{23} & i=1 \\
s_{123}-s_{12} & i=2 \\
s_{123}+p_1^2-s_{23}-{2p_1^2s_{123}\over s_{12}} & i=3 \\
p_1^2-s_{12} & i=4 \\
\end{array}
\right. .
\end{displaymath}

The scalar integral in $D=4-2\e$ can be written \cite{EGY,BDK1},
\begin{equation}
\Ihat_4[1] =
\frac{c_\Gamma}{\e^2}\left((\alpha_2\alpha_4)^\e + 2 (\alpha_1\alpha_3)^\e
-(\alpha_1\alpha_2)^\e
-(\alpha_1\alpha_4)^\e\right )
+ 2\Ld{0}(p_1,p_2,p_3) + {\cal O}(\e),
\end{equation}
where,
\begin{equation}
\Ld{0}(p_1,p_2,p_3) =
\Li2\left(1-\frac{\alpha_4}{\alpha_3}\right)
-\Li2\left(1-\frac{\alpha_3}{\alpha_2}\right)
+\frac{1}{2}
\log\left(\frac{\alpha_2\alpha_4}{\alpha_3^2}\right)
\log\left(\frac{\alpha_1}{\alpha_2}\right).
\end{equation}

In constructing the tensor integrals in $D=4-2\e$, we see
from equations~(\ref{eq:Inxi}--\ref{eq:Inxixjxkxl}) that the
box integral in higher dimensions is needed.
In fact, in $D=6-2\e$, the box integral is infrared and ultraviolet finite.
This can be seen by inspection of
equation~(\ref{eq:d-change}) and noting that the
pinchings with $m=1,2$ and 4 in the expression,
\begin{displaymath}
\Ihat_4[1] + \sum_{m=1}^4 \ghat_m\Ihat_3^{(m)}[1],
\end{displaymath}
are proportional to $1/\e^2$ and, when combined with the appropriate
$\ghat$ factor, precisely cancel with the pole structure of the box
integral.
The final pinching ($m=3$)  corresponds to the triangle
graph with three massive external legs which is itself finite.
Altogether, we find
that the adjacent box integral in $D=6$ is,
\begin{eqnarray}
\I_4^{D=6}[1]
&=&
-\frac{2s_{12}s_{23}}{\Delta_4}
\left(
\Ld{0}(p_1,p_2,p_3)
+ \frac{1}{2}\left(s_{123}+p_1^2-s_{23}
-\frac{2p_1^2s_{123}}{s_{12}}\right)
\Lc{0}(p_1,p_{23})\right) \nonumber \\
&\equiv& \Ld{1S}(p_1,p_2,p_3),
\label{eq:Ld1sadj}
\end{eqnarray}
where $\Lc{0}(p_1,p_{23}) = \I_3[1]$ is defined in equation~(\ref{eq:I3}).
Because of the finiteness properties of the three mass triangle,
we will find repeatedly that the $m=3$ pinching should be treated
differently from the other three.

\subsubsection{Scalar integrals in higher dimensions}

For higher dimensions, we just reuse equation~(\ref{eq:d-change}),
noting that the triangle
pinchings in $D = 6-2\e$ reintroduce ultraviolet poles.
These can easily be isolated by adding and subtracting combinations of
 scalar integrals as in section~3.1.2.
Explicitly,
\begin{eqnarray}
\I_4^{D=8-2\e}[1]
&=& \Ld{2S}(p_1,p_2,p_3) +\frac{c_\Gamma}{6}
\left(\frac{(-s_{123})^{-\e}}{\e}+ \frac{11}{3}\right),
\label{eq:I4adj8}
\\
\I_4^{D=10-2\e}[1]
&=& \Ld{3S}(p_1,p_2,p_3) +\frac{c_\Gamma (s_{123}+s_{12}+s_{23}+p_1^2)}{120}
\left(\frac{(-s_{123})^{-\e}}{\e}+ \frac{107}{30}\right),
\label{eq:I4adj10}
\\
\I_4^{D=12-2\e}[1]
&=& \Ld{4S}(p_1,p_2,p_3) +\frac{c_\Gamma P}{2520}
\left(\frac{(-s_{123})^{-\e}}{\e}+ \frac{129}{35}\right),
\label{eq:I4adj12}
\end{eqnarray}
where,
\begin{displaymath}
P = s_{123}^2+s_{12}^2+s_{23}^2+p_1^4
+s_{123}s_{12}+s_{123}s_{23}+s_{123}p_1^2
+s_{12}p_1^2+s_{23}p_1^2+\frac{s_{12}s_{23}}{2}.
\end{displaymath}
The finite parts of the higher dimension boxes are given by,
\begin{eqnarray}
\Ld{2S}(p_1,p_2,p_3) &=&
-\frac{s_{12}s_{23}}{3\Delta_4}
\Biggl(s_{12}s_{23}\Ld{1S}(p_1,p_2,p_3)
\nonumber \\&&
 +\left(s_{123}+p_1^2-s_{23}
-\frac{2p_1^2s_{123}}{s_{12}}\right)\Lc{1S}(p_1,p_{23})
\nonumber \\&&
 +\frac{s_{23}}{2}\log\left(\frac{s_{123}}{s_{23}}\right)
+s_{12}\log\left(\frac{s_{123}}{s_{12}}\right)
-\frac{p_1^2}{2}\log\left(\frac{s_{123}}{p_1^2}\right)\Biggr),
\label{eq:Ld2sadj}
\\
\Ld{3S}(p_1,p_2,p_3)&=&-\frac{s_{12}s_{23}}{5\Delta_4}
\Biggl(s_{12}s_{23}\Ld{2S}(p_1,p_2,p_3)
\nonumber \\&&
 +
\left(s_{123}+p_1^2-s_{23}
-\frac{2p_1^2s_{123}}{s_{12}}\right)
\Lc{2S}(p_1,p_{23})
\nonumber \\
&& +\frac{s_{23}^2}{24}\log\left(\frac{s_{123}}{s_{23}}\right)
+\frac{s_{12}^2}{12}\log\left(\frac{s_{123}}{s_{12}}\right)
   -\frac{p_1^4}{24}\log\left(\frac{s_{123}}{p_1^2}\right)
+\frac{s_{12}s_{23}}{12}\Biggr),
\label{eq:Ld3sadj}
\\
\Ld{4S}(p_1,p_2,p_3)&=&-\frac{s_{12}s_{23}}{7\Delta_4}
\Biggl(s_{12}s_{23}\Ld{3S}(p_1,p_2,p_3)
\nonumber \\&&
 +
\left(s_{123}+p_1^2-s_{23}
-\frac{2p_1^2s_{123}}{s_{12}}\right)
\Lc{3S}(p_1,p_{23})
\nonumber \\
&& +\frac{s_{23}^3}{360}\log\left(\frac{s_{123}}{s_{23}}\right)
+\frac{s_{12}^3}{180}\log\left(\frac{s_{123}}{s_{12}}\right)
   -\frac{p_1^6}{360}\log\left(\frac{s_{123}}{p_1^2}\right)\nonumber \\
&&
+\frac{s_{12}s_{23}(s_{123}+s_{12}+s_{23}+p_1^2)}{720}\Biggr).
\label{eq:Ld4sadj}
\end{eqnarray}
The $D=6-2\e$ and $D=8-2\e$ box integrals explicitly appear in the
momentum space tensor structure with one and two factors of
$g_{\mu\nu}$ respectively.
All of these integrals appear either directly or indirectly
in the tensor box integrals of equations~(\ref{eq:Inxi}--\ref{eq:Inxixjxkxl}).

\begin{figure}\vspace{8cm}
\includegraphics{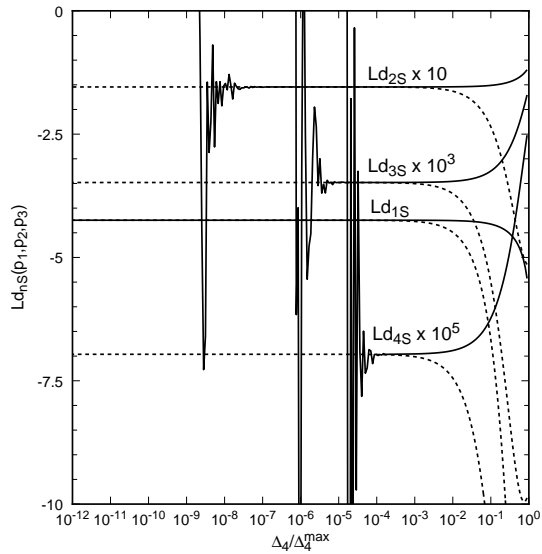}
\caption[]{The finite functions for the one mass box graph as a function of
$\Delta_4/\Delta_4^{{\rm max}}$ where
$\Delta_4^{{\rm max}} = 2 s_{12}s_{23}(s_{123}-s_{12}-s_{23})$.
The phase space point is $s_{123} = 1$, $s_{12} = 0.4$, $s_{23} = 0.08$
and $p_1^2$ altered so the limit is approached and the functions have been
evaluated in double precision Fortran.
The dashed lines show the
approximate form for the function in the limit $\Delta_4 \to 0$,
retaining only the first term of the
Taylor expansion as given in Appendix C.}
\label{fig:ldns}
\end{figure}

Once again, all of these functions
are well behaved as $\Delta_4 \to 0$ and group
a variety of dilogarithms, logarithms and  constants together
in a non-trivial way.  This is shown in Fig.~\ref{fig:ldns}
for a particular point in phase space; $s_{123} = 1$, $s_{12} = 0.4$,
$s_{23} = 0.08$
with $p_1^2$ varying so the $\Delta_4 \to 0$ limit is approached.
We see that although $\Ld{1S}$, with a single inverse power of the
Gram determinant,
is  numerically stable, the functions with more powers of Gram determinant
in the denominator break down at much larger values of $\Delta_4$.
In all cases, the function is well approximated by the first term
of the Taylor expansion provided $\Delta_4 < 10^{-4} \Delta_4^{{\rm max}}$.
These approximations are collected in Appendix C.

\subsubsection{Tensor integrals}

Armed with the integrals in $D > 4$, we return to the tensor integrals and use
equations~(\ref{eq:Inxi}--\ref{eq:Inxixjxkxl}) as the starting point.
Rewriting equation~(\ref{eq:Inxi}) for $n=4$, $D=4-2\e$ and noting that
we have eliminated $x_3$ so that $i=1,2$ and 4 only, we have,
\begin{displaymath}
\I_4[x_i] = -\g_i \I_4^{D=6}[1]
-  \sum_{m=1}^4 \eta_{im}\alpha_i\alpha_m \I_3^{(m)}[1].
\end{displaymath}
The factor $\alpha_i\alpha_m$ multiplying the triangle pinchings
will always produce a factor of $1/s$. For triangle graphs with
at least one massless leg (pinchings $m=1,2$ and 4),
the contribution is $\sim 1/\e^2$ and
 will combine with similar poles from other Feynman diagrams.
On the other hand, the $m=3$ pinching (corresponding to the
triangle graph  with momenta $p_{1}$ and $p_{23}$ flowing outwards)
is finite and,
provided the value of $s$ is related
kinematically to that triangle pinching, there may be a possibility
of cancellation with other triangle Feynman graphs.
However, for the case $i=1$ and $m=3$, the associated invariant mass
is $s_{12}$.
This term cannot combine with any other naturally generated
triangle graph with the same kinematics.
Therefore, we group this term with the box integral,
by adding and subtracting,
\begin{displaymath}
\sum_{m=1}^4 \eta_{im}\alpha_i\alpha_m \I_3^{(3)}[1]
= \g_i \I_3^{(3)}[1],
\end{displaymath}
so that,
\begin{equation}
\I_4[x_i] = \g_i \Ld{1}(p_1,p_2,p_3)
 -  \sum_{m=1}^4 \eta_{im}\alpha_i\alpha_m
\left(\I_3^{(m)}[1]-\I_3^{(3)}[1]\right),
\label{eq:I4axi}
\end{equation}
with,
\begin{equation}
\Ld{1}(p_1,p_2,p_3) = -\left(\I_4^{D=6}[1] +\I_3^{(3)}[1]\right).
\end{equation}
The only non-zero entries in $\eta_{im}$ for $i\neq 3$ and $m \neq 3$ are
$i = 2,~m=4$ or $i=4,~m=2$ corresponding to
$\alpha_i\alpha_m = -1/s_{23}$ which {\em is} appropriate for $\I_3^{(3)}[1]$.

For integrals with more Feynman parameters it is
convenient to introduce the following functions,
\begin{equation}
\Ld{ni_1\ldots i_{n-1}}(p_1,p_2,p_3) = -
\biggl\{n\I_4^{D=6}[x_{i_1}\ldots x_{i_{n-1}}]
+\I_3^{(3)}[x_{i_1}\ldots x_{i_{n-1}}]\biggr\},
\end{equation}
for $n=2,3$ and 4.
Suppressing the arguments of $\Ld{n}$, and using
equations~(\ref{eq:Inxixj}-\ref{eq:Inxixjxkxl}), we find,
\begin{eqnarray}
\label{eq:I4axixj}
\I_4[x_ix_j]
&=&
\g_i \Ld{2j} -
\eta_{ij}\alpha_i\alpha_j \biggl\{ \Ld{1}+\I_3^{(3)}[1]\biggr\}
 -  \sum_{m=1}^4 \eta_{im}\alpha_i\alpha_m
\biggl\{\I_3^{(m)}[x_j]-\I_3^{(3)}[x_j]\biggr\},
\\
\label{eq:I4axixjxk}
\I_4[x_ix_jx_k]
&=&
\g_i \Ld{3jk}
- \frac{1}{2}\eta_{ij}\alpha_i\alpha_j \biggl\{\Ld{2k}+\I_3^{(3)}[x_k]\biggr\}
- \frac{1}{2}\eta_{ik}\alpha_i\alpha_k \biggl\{\Ld{2j}+\I_3^{(3)}[x_j]\biggr\}
\nonumber\\
&&\quad
 -  \sum_{m=1}^4 \eta_{im}\alpha_i\alpha_m
\biggl\{\I_3^{(m)}[x_jx_k]-\I_3^{(3)}[x_jx_k]\biggr\},
\\
\label{eq:I4axixjxkxl}
\I_4[x_ix_jx_kx_l]
&=&
\g_i \Ld{4jkl}
- \frac{1}{3}\eta_{ij}\alpha_i\alpha_j
\biggl\{\Ld{3kl}+\I_3^{(3)}[x_kx_l]\biggr\}
- \frac{1}{3}\eta_{ik}\alpha_i\alpha_k
\biggl\{\Ld{3jl}+\I_3^{(3)}[x_jx_l]\biggr\}
\nonumber \\
&&
- \frac{1}{3}\eta_{il}\alpha_i\alpha_l
\biggl\{\Ld{3jk}+\I_3^{(3)}[x_jx_k]\biggr\}
 -  \sum_{m=1}^4 \eta_{im}\alpha_i\alpha_m
\biggl\{\I_3^{(m)}[x_jx_kx_l]-\I_3^{(3)}[x_jx_kx_l]\biggr\}.
\nonumber \\
\end{eqnarray}
Since we have systematically eliminated $x_3$ using the delta function,
$i,~j,~k$ and $l$ run over $1,~2$ and 4.   This guarantees that
the coefficients of the form $\eta_{ij}\alpha_i\alpha_j$  are only non-zero
for $i=2$ and $j=4$ (or vice-versa).
In these cases, $\eta_{24}\alpha_2\alpha_4 =  -1/s_{23}$,
which is
again appropriate for the $m=3$  pinching to form a completely massive
triangle,
$\I_3^{(3)}[1]$.
Altogether, equations~(\ref{eq:I4axi},\ref{eq:I4axixj}--\ref{eq:I4axixjxkxl})
are sufficient
to completely describe the tensor structure of the adjacent box.

However, in order to determine the $\Ld{ni_1\ldots i_{n-1}}$ combinations, we
need tensor integrals for $D=6-2\e$ dimensional box graphs with
two or more Feynman parameters.
These can be obtained from equation~(\ref{eq:Inxixj}) once the $D=6-2\e$
box integral with a single Feynman parameter, $\I_4^{D=6-2\e}[x_i]$
is known. This can be derived by differentiating (\ref{eq:deriv})
which indicates that $\I_4^{D=6-2\e}[x_i]$
is also finite as $\e \to 0$.
To see this,
we reuse equation~(\ref{eq:Inxi}) and our usual trick of adding and
subtracting combinations of the $m=3$ pinching in $D=6-2\e$,
\begin{eqnarray}
\I_4^{D=6-2\e}[x_i] &=&
\g_i \left(-(3-2\e)\I_4^{D=8-2\e}[1] - \I_3^{D=6-2\e (3)}[1]\right)\nonumber \\
&&\qquad
-\sum_{m=1}^4 \eta_{im}\alpha_i\alpha_m
\left(\I_3^{D=6-2\e (m)}[1]- \I_3^{D=6-2\e (3)}[1] \right).\nonumber
\end{eqnarray}
Both brackets are separately finite.   First, the divergent part of
$\I_4^{D=8-2\e}[1]$ precisely cancels against that of $\I_3^{D=6-2\e (3)}[1]$.
Second, all triangles in $D=6-2\e$ dimensions have $1/\e$ poles and the
difference of any two, is either zero or a log.
Further differentiation does not change the finiteness properties of
the $\I_4^{D=6-2\e}$ tensor integrals.

The $\Ld{ni_1\ldots i_{n-1}}$ combinations are also well behaved in
certain kinematic limits.   For example,
$\I_4^{D=6}[1]$ and $\I_3^{(3)}[1]$ are finite as $s_{12}\to 0$ or $s_{23} \to
0$.
Just as differentiating functions which are
finite as $\e \to 0$ does not introduce
poles in $\e$, neither can it introduce poles in
the kinematic invariants $s_{12}$ or $s_{23}$.
As an example, consider the function $\Ld{24}$ given by,
\begin{eqnarray}
\Ld{24}(p_1,p_2,p_3) &=&  \frac{2(p_1^2-s_{12})}{s_{12}s_{23}} \left(
3\Ld{2S}(p_1,p_2,p_3) + \Lc{1S}(p_1,p_{23}) \right) - \frac{s_{12}}{s_{23}}
\Lc{1}(p_3,p_{12})\nonumber \\
&&\qquad + \frac{p_1^2}{s_{23}} \Lc{1}(p_{23},p_1),\nonumber
\end{eqnarray}
which appears to contain a pole in $s_{12}$.
In the $s_{12}\to 0$ limit, $\Delta_4 \to -2s_{23}s_{123}p_1^2$ and
\begin{displaymath}
\Ld{2S} \to \frac{2s_{23}s_{123}p_1^2}{3\Delta_4}\Lc{1S}(p_1,p_{23})
\to -\frac{1}{3}\Lc{1S}(p_1,p_{23}),
\end{displaymath}
so that,
\begin{displaymath}
\lim_{s_{12}\to 0} \left (3\Ld{2S}(p_1,p_2,p_3) + \Lc{1S}(p_1,p_{23}) \right)
\to 0,
\end{displaymath}
and therefore,
\begin{displaymath}
\lim_{s_{12}\to 0} s_{12} \times \Ld{24}(p_1,p_2,p_3) \to 0.
\end{displaymath}
Similarly, $\Ld{24}$ contains no power-like divergences\footnote{
Although these functions do not behave as inverse powers of the vanishing
kinematic variables, they do
contain logarithms of $s_{12}$ and $s_{23}$.
This is because the $\e \to 0$ limit has already been taken,
and the order of taking the two limits does not commute.
For next-to-leading order calculations,
we only approach the limit $s_{ij} \to 0$
and $\e$ can safely be taken to zero first.}
in the $s_{23} \to 0$ limit
and, with a little more work, it can be shown that,
\begin{displaymath}
\lim_{s_{23}\to 0} s_{23} \times \Ld{24}(p_1,p_2,p_3) \to 0.
\end{displaymath}
Once again, these functions combine dilogarithms, logarithms and constants
in a highly non-trivial way to form well behaved building blocks.
Explicit forms for the $\Ld{ni_1\ldots i_{n-1}}$ functions for $n=1,2$ and 3
are given in Appendix B.

\subsection{The one-mass box}

The higher dimension and Feynman parameter integrals for the one-mass box
integral obtained by taking
$p_1^2 \to 0$ can also be constructed in a similar way.
For this kinematic configuration, the $\alpha$ parameters are defined by,
\begin{displaymath}
\alpha_1\alpha_4 \, s_{123}=-1, \quad
\alpha_1\alpha_3 \, s_{12}=-1,\quad
\alpha_2\alpha_4 \, s_{23}=-1,
\end{displaymath}
while,
\begin{displaymath}
\begin{array}{lr}
\eta_{ij} =
\left(
\begin{array}{rrrr}
0 & 0 & 1 & 0 \\ 0 & 0 & -1 & 1 \\
1 & -1 & 0 & 0 \\ 0 & 1 & 0 & 0
\end{array}
\right),
& N_4  = \frac{1}{2},
\end{array}
\end{displaymath}
and,
\begin{displaymath}
\Delta_4^{1m} = 2s_{12}s_{23}\left(s_{123}-s_{12}-s_{23}\right).
\end{displaymath}
In terms of invariants,
\begin{displaymath}
\frac{\gamma_i}{  \prod_{j=1}^4 \alpha_j} =
\left\{
\begin{array}{ll}
-s_{23} & i=1 \\
s_{123}-s_{12} & i=2 \\
s_{123}-s_{23} & i=3 \\
-s_{12} & i=4 \\
\end{array}
\right. .
\end{displaymath}
The scalar integral can be written,
\begin{equation}
\Ihat_4^{1m}[1] =
\frac{c_\Gamma}{\e^2}\left(2(\alpha_1\alpha_3)^\e + 2 (\alpha_2\alpha_4)^\e
-2(\alpha_1\alpha_4)^\e\right )
+ 2\Ld{0}^{1m}(p_1,p_2,p_3) + {\cal O}(\e),
\end{equation}
where,
\begin{equation}
\Ld{0}^{1m}(p_1,p_2,p_3) =
\Li2\left(1-\frac{\alpha_4}{\alpha_3}\right)
+\Li2\left(1-\frac{\alpha_1}{\alpha_2}\right)
+\log\left(\frac{\alpha_4}{\alpha_3}\right)
\log\left(\frac{\alpha_1}{\alpha_2}\right)
-\frac{\pi^2}{6}.
\end{equation}
As expected, in $D=6-2\e$ dimensions, the scalar integral is finite,
\begin{equation}
\I_4^{1m,D=6}[1] =
-\frac{2s_{12}s_{23}}{\Delta_4^{1m}}\Ld{0}^{1m}(p_1,p_2,p_3) \equiv
\Ld{1S}^{1m}(p_1,p_2,p_3).
\label{eq:Ld1s1m}
\end{equation}
In higher dimensions, the scalar integrals satisfy analagous equations to
(\ref{eq:I4adj8}--\ref{eq:I4adj12})  with \linebreak $p_1^2 \to 0$,
and finite parts given by,
\begin{eqnarray}
\Ld{2S}^{1m}(p_1,p_2,p_3) &=&
-\frac{s_{12}s_{23}}{3\Delta_4^{1m}}
\Biggl(s_{12}s_{23}\Ld{1S}^{1m}(p_1,p_2,p_3)
+s_{23}\log\left(\frac{s_{123}}{s_{23}}\right)
  +s_{12}\log\left(\frac{s_{123}}{s_{12}}\right)\Biggr),
\label{eq:Ld2s1m}
\nonumber \\
&& \\
\Ld{3S}^{1m}(p_1,p_2,p_3)&=&-\frac{s_{12}s_{23}}{5\Delta_4^{1m}}
\Biggl(s_{12}s_{23}\Ld{2S}^{1m}(p_1,p_2,p_3)
+\frac{s_{23}^2}{12}\log\left(\frac{s_{123}}{s_{23}}\right)
+\frac{s_{12}^2}{12}\log\left(\frac{s_{123}}{s_{12}}\right)
\nonumber \\
&&\qquad+\frac{s_{12}s_{23}}{12}\Biggr),
\label{eq:Ld3s1m}
\\
\Ld{4S}^{1m}(p_1,p_2,p_3)&=&-\frac{s_{12}s_{23}}{7\Delta_4^{1m}}
\Biggl(s_{12}s_{23}\Ld{3S}^{1m}(p_1,p_2,p_3)
+\frac{s_{23}^3}{180}\log\left(\frac{s_{123}}{s_{23}}\right)
+\frac{s_{12}^3}{180}\log\left(\frac{s_{123}}{s_{12}}\right)
\nonumber \\
&&\qquad+\frac{s_{12}s_{23}(s_{123}+s_{12}+s_{23})}{720}\Biggr).
\label{eq:Ld4s1m}
\end{eqnarray}

\begin{figure}[t]\vspace{8cm}
\includegraphics{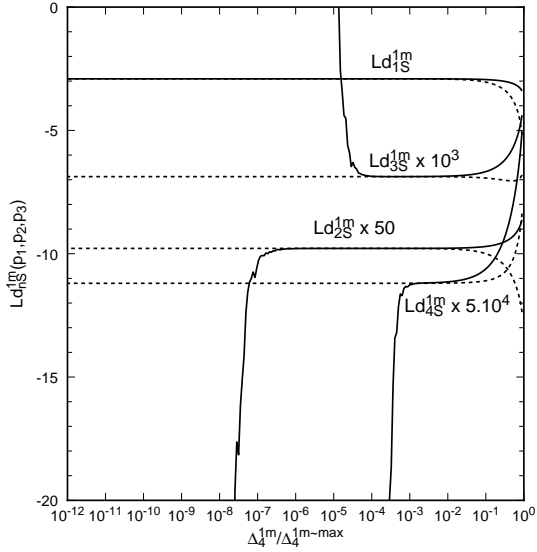}
\caption[]{The finite functions for the box graph with two adjacent
massive legs as a function of
$\Delta_4^{1m}/\Delta_4^{1m~{\rm max}}$.
The functions have been
evaluated in double precision Fortran.
The dashed lines show the
approximate form for the function in the limit $\Delta_4^{1m} \to 0$,
retaining only the first term of the
Taylor expansion as given in Appendix C.}
\label{fig:ldns1m}
\end{figure}

The stability of these functions as $\Delta_4^{1m} \to 0$ is
illustrated in fig.~\ref{fig:ldns1m}
for a particular point in phase space; $s_{123} = 1$, $s_{12} = 0.3$,
with $s_{23}$ varying so the limit is approached.
The maximum
possible value of $\Delta_4^{1m}$ occurs when
$s_{23} = (s_{123}-s_{12})/2$;
i.e.~$\Delta_4^{1m~{\rm max}} = s_{12}(s_{123}-s_{12})^2/2$.
As before,  $\Ld{1S}^{1m}$, with a single inverse power of the
Gram determinant,
is  numerically stable.
However there are numerical instabilities for
the other functions with more powers of Gram determinant
in the denominator.
In all cases, the function is well approximated by the first term
of the Taylor expansion provided $\Delta_4^{1m} < 10^{-3}
\Delta_4^{1m~{\rm max}}$.

Unlike the $p_1^2 \neq 0$ case, the scalar triangle pinchings all contain
infrared poles and there is no benefit in absorbing the $m=3$ piece in the
tensor integrals.
Therefore we introduce,
\begin{equation}
\Ld{ni_1\ldots i_{n-1}}^{1m}(p_1,p_2,p_3) = -
n\I_4^{1m,D=6}[x_{i_1}\ldots x_{i_{n-1}}],
\end{equation}
for $n=1,~2,~3$ and 4. Using equations~(\ref{eq:Inxi}--\ref{eq:Inxixjxkxl})
with $n=4$ and $D=4-2\e$ we find,
\begin{eqnarray}
\I_4^{1m}[x_i] &=& \g_i \Ld{1}^{1m}
 -  \sum_{m=1}^4 \eta_{im}\alpha_i\alpha_m \I_3^{(m)}[1],\\
\I_4^{1m}[x_ix_j]
&=&
\g_i \Ld{2j}^{1m} -
\eta_{ij}\alpha_i\alpha_j  \Ld{1}^{1m}
 -  \sum_{m=1}^4 \eta_{im}\alpha_i\alpha_m \I_3^{(m)}[x_j],\\
\I_4^{1m}[x_ix_jx_k]
&=&
\g_i \Ld{3jk}^{1m}
-\frac{1}{2} \eta_{ij}\alpha_i\alpha_j \Ld{2k}^{1m}
-\frac{1}{2} \eta_{ik}\alpha_i\alpha_k \Ld{2j}^{1m}\nonumber \\
&&\qquad
 -  \sum_{m=1}^4 \eta_{im}\alpha_i\alpha_m \I_3^{(m)}[x_jx_k],\\
\I_4^{1m}[x_ix_jx_kx_l]
&=&
\g_i \Ld{4jkl}^{1m}
-\frac{1}{3} \eta_{ij}\alpha_i\alpha_j \Ld{3kl}^{1m}
-\frac{1}{3} \eta_{ik}\alpha_i\alpha_k \Ld{3jl}^{1m}
-\frac{1}{3} \eta_{il}\alpha_i\alpha_l \Ld{3jk}^{1m}\nonumber \\
&&\qquad
 -  \sum_{m=1}^4 \eta_{im}\alpha_i\alpha_m
\I_3^{(m)}[x_jx_kx_l].
\end{eqnarray}

As in the previous section, the
$\Ld{ni_1\ldots i_{n-1}}^{1m}$ functions
are finite as $\e \to 0$ and contain no power-like divergences
in
the $s_{12} \to 0$, $s_{23}\to 0$ and $\Delta_4^{1m} \to 0$ limits.
For convenience,
explicit forms are given in Appendix~B.

\subsection{The opposite two-mass box}

The two mass box graph where the massive legs sit on opposite
sides is a special case because the scalar integral itself is not finite
as $\Delta_4^{opp}\to 0$.   We must therefore proceed
with care.
To make best use of the symmetry under $p_1 \leftrightarrow p_3$,
it is convenient to write,
\begin{displaymath}
\P^\mu = -(1-x_1)p_1^\mu - (x_3+x_4)p_2^\mu - x_4 p_3^\mu.
\end{displaymath}
Under this flip symmetry,
$x_1 \leftrightarrow x_4$, $x_2 \leftrightarrow x_3$ and
\begin{displaymath}
\P^\mu \to -\P^\mu - p_1^\mu-p_2^\mu-p_3^\mu.
\end{displaymath}

In this kinematic configuration, the $\alpha$ parameters can be defined by,
\begin{displaymath}
\begin{array}{rr}
\alpha_1\alpha_4 \, s_{123}=-\lambda, & \alpha_2\alpha_3 \, p_2^2 =-1 \\
  \alpha_1\alpha_3 \, s_{12}=-1, & \alpha_2\alpha_4 \, s_{23}=-1,
\end{array}
\end{displaymath}
where $\lambda$ is an extra kinematic variable that ensures that the
$\alpha_i$ are independent,
\begin{displaymath}
\lambda = \frac{s_{123}p_2^2}{s_{12}s_{23}}.
\end{displaymath}
With this choice of $\alpha_i$,
\begin{displaymath}
\eta_{ij} = (1-\lambda)
\left(
\begin{array}{rrrr}
0 & 0 & 1 & -1 \\ 0 & 0 & -\lambda & 1 \\
1 & -\lambda & 0 & 0 \\ -1 & 1 & 0 & 0
\end{array}\right),
\qquad
 N_4  = \frac{1}{2}(1-\lambda)^2.
\end{displaymath}
Each row of $\eta$ naturally couples together two of the pinchings (triangles)
of this box, so we might expect such structure to dominate the integrals.
The associated Gram determinant is given by,
\begin{displaymath}
\Delta_4^{opp}    = 2(s_{12}s_{23}-p_2^2s_{123})s_{13}
		 = \frac{2(1-\lambda)s_{13}}{\prod_{i=1}^4 \alpha_i},
\end{displaymath}
where,
$$
s_{13} = s_{123}-s_{12}-s_{23}+p_2^2,
$$
and,
\begin{displaymath}
\frac{\gamma_i}{  \prod_{j=1}^4 \alpha_j} = (1-\lambda)
\left\{
\begin{array}{ll}
p_2^2-s_{23} & i=1 \\
s_{123}-s_{12} & i=2 \\
s_{123}-s_{23} & i=3 \\
p_2^2-s_{12} & i=4 \\
\end{array}
\right. .
\end{displaymath}
We note that the presence of the Gram determinant is synonymous with a
factor of $(1-\lambda)$.

The scalar integral for the opposite box in $D=4-2\e$ is given by,
\begin{equation}
\Ihat_4^{opp}[1]
=
\frac{2}{(1-\lambda)}
\left (
\frac{c_\Gamma}{\e^2}
\left (
(\alpha_1\alpha_3)^\e + (\alpha_2\alpha_4)^\e
- (\alpha_2\alpha_3)^\e - (\alpha_1\alpha_4)^\e\lambda^{-\e}\right)
+\Ld{0}^{opp}(p_1,p_2,p_3) \right) + {\cal O}(\e),
\end{equation}
where the finite part $\Ld{0}$ can be written,
\begin{eqnarray}
\Ld{0}^{opp}(p_1,p_2,p_3) &=&
\Li2\left(1-\lambda\right)
+\Li2\left(1-\frac{\alpha_4}{\lambda \alpha_3}\right)
+\Li2\left(1-\frac{\alpha_1}{\lambda \alpha_2}\right)\nonumber \\
&&\quad
-\Li2\left(1-\frac{\alpha_4}{\alpha_3}\right)
-\Li2\left(1-\frac{\alpha_1}{\alpha_2}\right)
+\log\left(\frac{\alpha_4}{\lambda \alpha_3}\right)
 \log\left(\frac{\alpha_1}{\lambda \alpha_2}\right).
\end{eqnarray}
As $\lambda \to 1$, there is a manifest singularity in $\Ihat_4^{opp}[1]$
since,
\begin{displaymath}
\Ld{0}^{opp}(p_1,p_2,p_3)  \to -
\log\left(\frac{\alpha_1}{\alpha_2}\right)
\log\left(\frac{\alpha_3}{\alpha_4}\right).
\end{displaymath}
This double logarithm can never combine with lower point scalar integrals to
form a combination well behaved as
$\Delta_4^{opp}\to 0$.  In fact, it is easy to see from
fig.~\ref{fig:boxin} that the only scalar integrals which are available by
pinching are the triangle integrals with one and two massive legs.   These
are pure poles in $\e$ and cannot be combined with the finite parts of the
opposite
box integral.  There is no appropriate function which can generate the
double logarithm as $\lambda \to 1$ and consequently no finite function
can be formed.
Since the matrix elements are in general finite in the limit of vanishing
Gram determinants, all occurences
of  $\Ld{0}^{opp}$ divided by the determinant must vanish.

On the other hand, in $D=6-2\e$, the opposite box is not only
finite as $\e \to 0$
as expected, but also as $\lambda \to 1$.
This is because $N_4$ is effectively $(\Delta^{opp}_4)^2$ and
its presence in the numerator of equation~(\ref{eq:d-change}) removes the
Gram determinant from the denominator.
Consequently, we see that,
\begin{equation}
\I_4^{opp~D=6-2\e}[1]
=  -\frac{1}{s_{13}}
\Ld{0}^{opp}(p_1,p_2,p_3),
\end{equation}
which, since,
\begin{eqnarray}
\lim_{s_{13} \to 0} \Ld{0}(p_1,p_2,p_3) &=&
\frac{s_{13}}{s_{12}s_{23}-p_2^2s_{123}}\left(
s_{12} \log \left( \frac{s_{123}}{s_{12}} \right)
 +s_{23} \log \left( \frac{s_{123}}{s_{23}} \right)
 - p_2^2 \log \left( \frac{s_{123}}{p_2^2} \right) \right)\nonumber \\
&&+ {\cal O}(s_{13}^2),
\label{eq:ldopplim}
\end{eqnarray}
is also finite as $s_{13} \to 0$.

In dealing with the all-massive triangle and adjacent box in
the previous sections, we have constructed groups of scalar integrals that
are finite both as $\e \to 0$ and $\Delta \to 0$.
For the opposite box, this is not easy to do since the
raw scalar integral is not finite in either of these limits.
Therefore, we follow the approach used for the two-mass triangle graph
and obtain the tensor integrals by
direct differentiation.
However, we note that the factor $1/(n-D)$ associated
with the differentiation formula (\ref{eq:deriv}) is itself divergent,
implying that the ${\cal O}(\e)$ term in $\Ihat_4^{opp}[1]$ is needed -
the so called $\e$-barrier.
We circumvent this, by using equation~(\ref{eq:Inxi}), to construct the
tensor integral with a single Feynman parameter.
Explicitly, we find,
\begin{eqnarray}
\Ihat_4^{opp}[a_3] &=& \frac{1}{(1-\lambda)}
\Biggl \{\frac{(\alpha_1-\lambda\alpha_2)\Ld{0}^{opp}(p_1,p_2,p_3)}
{\alpha_1\alpha_3+\alpha_2\alpha_4-\alpha_1\alpha_4-\lambda\alpha_2\alpha_3}
\nonumber \\
&&\qquad
-\frac{c_\Gamma}{\e^2}
\left(\frac{(\alpha_2\alpha_4)^{\e}
- (\alpha_2\alpha_3)^{\e}}{\alpha_4-\alpha_3}\right)
+\frac{\lambda c_\Gamma}{\e^2}
\left(\frac{(\alpha_1\alpha_4)^{\e}\lambda^{-\e}
- (\alpha_1\alpha_3)^{\e}}{\alpha_4-\lambda\alpha_3}\right) \Biggr \},\\
\Ihat_4^{opp}[a_4] &=& \frac{1}{(1-\lambda)}
\Biggl \{\frac{(\alpha_2-\alpha_1)\Ld{0}^{opp}(p_1,p_2,p_3)}
{\alpha_1\alpha_3+\alpha_2\alpha_4-\alpha_1\alpha_4-\lambda\alpha_2\alpha_3}
\nonumber \\
&&\qquad
+\frac{c_\Gamma}{\e^2}
\left(\frac{(\alpha_2\alpha_4)^{\e}
- (\alpha_2\alpha_3)^{\e}}{\alpha_4-\alpha_3}\right)
-\frac{c_\Gamma}{\e^2}
\left(\frac{(\alpha_1\alpha_4)^{\e}\lambda^{-\e}
- (\alpha_1\alpha_3)^{\e}}{\alpha_4-\lambda\alpha_3}\right) \Biggr \}.
\end{eqnarray}
This is enough to construct all the necessary integrals, since
those for $a_2$ and $a_1$ can be obtained by the simultaneous exchanges,
$a_1 \leftrightarrow a_4$,
$a_2\leftrightarrow a_3$
(and $p_1 \leftrightarrow p_3$),
while integrals with higher powers can be obtained by direct differentiation.
The only subtlety is in differentiating $\Ld{0}$, where it is useful
to re-express some of the logarithms as single poles in $\e$.
For example,
$$
\frac{\partial \Ld{0}(p_1,p_2,p_3)}{\partial \alpha_4}
=   \frac{\log\left(\frac{\alpha_4}{\alpha_3}\right)}{\alpha_4-\alpha_3}
-   \frac{\log\left(\frac{\alpha_4}{\lambda\alpha_3}\right)}
{\alpha_4-\lambda\alpha_3}
+ \frac{c_\Gamma}{\e}\left(\frac{(\alpha_1\alpha_4)^{\e}\lambda^{-\e}
-(\alpha_2\alpha_4)^{\e}}{\alpha_4}\right) + {\cal O}(\e).
$$
This helps to ensure that although the double poles in $\e$ are
still divided by $(1-\lambda)$, the single poles are finite as $\lambda \to 1$.
In fact, when the tensor integral is multipled by
a factor of $\e$ we also expect that the resulting single logarithms occur in
groups that are finite as $\lambda \to 1$.
We therefore introduce the auxiliary
functions,
\begin{eqnarray}
\Lcd{n}(p_1,p_2,p_3) &=&
-  \lim_{\e \to 0}  \left( \e \times \I_4^{opp}[x_4^{n}] \right )\nonumber \\
&=& - \frac{1}{(1-\lambda)} \lim_{\e \to 0} \left(
\alpha_4\alpha_1 \I_3^{(1)}[x_4^{n-1}] -
\alpha_4\alpha_2 \I_3^{(2)}[x_4^{n-1}]  \right )\nonumber \\
&=& \frac{2s_{13}}{\Delta_4^{opp}}
\left(  s_{12} \Lc{n}^{2m}(p_{12},p_3) - p_2^2 \Lc{n}^{2m}(p_2,p_3) \right ),
\end{eqnarray}
for $n=1,\ldots,4$ and,
\begin{equation}
\Lcd{0}(p_1,p_2,p_3) = \frac{2s_{13}}{\Delta_4^{opp}}\log\left(
\frac{s_{12}s_{23}}{p_2^2s_{123}}\right).
\end{equation}
Because these functions contain only a
single power of the Gram determinant,
they are not difficult to evaluate numerically.

The tensor integrals are straightforward to derive,
but can be rather lengthy,  for example,
\begin{eqnarray}
\I_4^{opp}[x_4^2] & = &
-\frac{(\prod_i \alpha_i)\alpha_4^2}{1-2\e}\frac{\partial}{\partial\alpha_4}
\Ihat^{opp}_4[a_4]\nonumber \\
&=&
\frac{c_\Gamma}{\e^2}
\frac{2s_{13}}{\Delta_4^{opp}} \Biggl(
 \frac{s_{12}^2 \left( (-s_{12})^{-\e}-(-s_{123})^{-\e} \right)}{(s_{123}
 -s_{12})^2} - \frac{p_2^4 \left( (-p_2^2)^{-\e}-(-s_{23})^{-\e}
 \right)}{(s_{23} - p_2^2)^2} \Biggr) \nonumber \\ & & \qquad
 + \frac{c_\Gamma}{\e} \frac{1}{s_{13}} \Biggl(
\frac{(-s_{123})^{-\e}}{(s_{123}
 -s_{12})} - \frac{(-s_{23})^{-\e}}{(s_{23}-p_2^2)} \Biggr) +
 \frac{2(s_{12}-p_2^2)^2}{s_{13}\Delta_4^{opp}} \Ld{0}(p_1,p_2,p_3)
 \nonumber \\ & & \qquad - \frac{(s_{12}-p_2^2)}{s_{13}}
 \Lcd{1}(p_1,p_2,p_3) - 2 \Lcd{2}(p_1,p_2,p_3).
\end{eqnarray}
Most of these terms are poles in $\e$.   For physical processes,
the $1/\e^2$ poles from different tensor integrals must
combine in such a way that the $\Delta^{opp}_4$ cancels, while the $1/\e$ poles
do not depend on the determinant
and are ready to cancel with contributions from other loop configurations.
In all cases, the scalar box function $\Ld{0}$ is associated with a single
inverse power of $\Delta_4^{opp}$.
As noted earlier,  for physical processes that are
finite as $\Delta_4^{opp}\to 0$, there must be cancellations amongst the
various tensor integrals so that no terms containing
$\Ld{0}/\Delta_4^{opp}$ remain.
We therefore choose to leave the $\Ld{0}$ functions exposed
to facilitate the cancellation of these terms.

For third and fourth rank tensor integrals, we also need to know
the $D=6-2\e$ box with one or two Feynman parameters in the numerator and the
$D=8-2\e$ box.
The former are finite as $\e \to 0$ and
are easily obtained by direct differentiation of $\I_4^{opp~D=6-2\e}[1]$
so that, for example,
\begin{eqnarray}
\I_4^{opp~D=6-2\e}[x_4] &=& \frac{1}{2s_{13}}\Biggl\{
\frac{(s_{12}-p_2^2)}{s_{13}} \Ld{0}(p_1,p_2,p_3)
 -  s_{12} \Lc{1}^{2m}(p_{12},p_3)
\nonumber \\
&&\qquad + p_2^2 \Lc{1}^{2m}(p_2,p_3)
-\log\left(\frac{s_{123}}{s_{23}}\right)
\Biggr \},
\end{eqnarray}
while the $D=8-2\e$ box is established using the dimension changing equation
(\ref{eq:d-change}),
\begin{eqnarray}
\I_4^{opp~D=8-2\e}[1] & = & \frac{1}{6s_{13}}
\Biggl \{
\frac{s_{12}s_{23}-p_2^2s_{123}}{s_{13}} \Ld{0}(p_1,p_2,p_3)
 - s_{12} \log \left( \frac{s_{123}}{s_{12}} \right)
- s_{23} \log \left( \frac{s_{123}}{s_{23}} \right)
 \nonumber \\ & & \qquad
 + p_2^2 \log \left( \frac{s_{123}}{p_2^2} \right)\Biggr \}
 + \frac{1}{6}
\left( \frac{(-s_{123})^{-\e}}{\e} + \frac{11}{3} \right)c_\Gamma .
\end{eqnarray}
As might be expected, both of these are finite as $s_{13} \to 0$ and this
can be seen directly from equation~(\ref{eq:ldopplim}).

In summary, the situation for box integrals is very similar to that for
triangle graphs.
When the scalar integral is finite as $\Delta_4 \to 0$, differentiating
- or equivalently adding factors of Feynman parameters -
does not introduce kinematic singularities.
Hence natural groupings of box and triangle integrals arise that are
finite as $\Delta_4 \to 0$.
Furthermore, the infrared and ultraviolet singularities can be
isolated easily.
Although we have explicitly worked through a subset of kinematic
configurations relevant to certain QCD processes, this method
is systematic and can be applied to processes with more general
kinematics (and particularly non-zero internal masses).

\section{Five point integrals}
\setcounter{equation}{0}

\begin{figure}\vspace{8cm}
\includegraphics{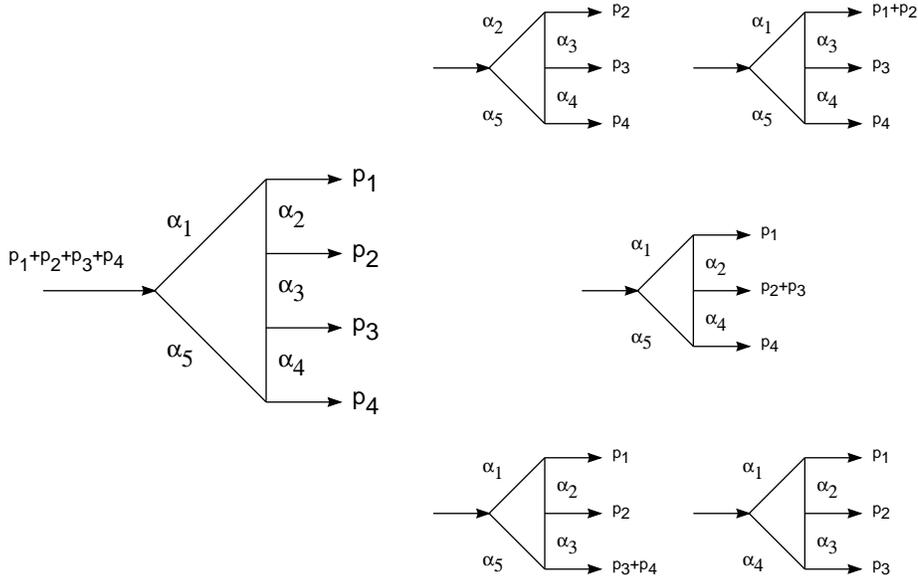}
\caption[]{The pentagon graph and each of
the five pinchings obtained by omitting
the internal line associated with $\alpha_m$ for $m=1,2,3,4$~and~5.
For box pinchings, the entering momentum is fixed by
momentum conservation.}
\label{fig:pentin}
\end{figure}

In this section we consider five point integrals with only one external mass.
The outflowing lightlike momenta are denoted $p_i$, $i=1,\ldots,4$
while the fifth  $p_5 = -p_{1234}$ is massive, $p_5^2 \neq 0$ as shown in
fig.~\ref{fig:pentin}.
The auxiliary momentum is then,
\begin{displaymath}
\P^\mu = -(1-x_1)p_1^\mu-(1-x_1-x_2)p_2^\mu - (x_4+x_5) p_3^\mu- x_5 p_4^\mu.
\end{displaymath}
We can make the choice,
\begin{displaymath}
\begin{array}{rr}
\alpha_1\alpha_5 \, s_{1234}=-\lambda, & \alpha_2\alpha_4 \, s_{23} =-1 \\
  \alpha_1\alpha_3 \, s_{12}=-1, & \alpha_2\alpha_5 \, s_{234}=-1 \\
  \alpha_1\alpha_4 \, s_{123}=-1, & \alpha_3\alpha_5 \, s_{34}=-1,
\end{array}
\end{displaymath}
with,
\begin{displaymath}
(1-\lambda)=\frac{1}{\alpha_3}\prod_{i=1}^5 \alpha_i
(s_{123}s_{234}-s_{23}s_{1234}).
\end{displaymath}
As in the opposite box integral, $\lambda$ is an extra kinematic variable
that ensures the $\alpha_i$ are independent.
It is the same variable that occurs in the third pinching which forms
an opposite box configuration.
The matrix $\eta_{ij}$ is given by,
\begin{displaymath}
\eta_{ij} =
\left(
\begin{array}{ccccc}
1 & -1 & 1-\lambda & 1 & -1 \\ -1 & 1 & \lambda-1 & 1-2\lambda & 1 \\
1-\lambda & \lambda-1 & (1-\lambda)^2 & \lambda-1 & 1-\lambda \\
1 & 1-2\lambda & \lambda-1 & 1 & -1 \\ -1 & 1 & 1-\lambda & -1 & 1
\end{array}
\right),
\end{displaymath}
and the normalisation factor is,
\begin{displaymath}
N_5=1-\lambda.
\end{displaymath}
The $\gamma_i$ are rather lengthy, but can be read off from $\eta_{ij}$.
The scalar pentagon integral is by now
well-known in $D=4$ \cite{5in4,HV,OV} and in $D=4-2\e$ \cite{EGY,BDK1,BDK2}
and can be written in terms of these
variables as,
\begin{equation}
\Ihat_5[1] = -\frac{1}{2N_5}
\sum_{m=1}^5 \ghat_m \Ihat_4^{(m)}[1]
  + \cal{O}(\e) .
\end{equation}
The five pinchings and the momenta associated with each is illustrated in
fig.~\ref{fig:pentin}.
The limit $N_5 \rightarrow 0$ corresponds to the vanishing of the Gram
determinant associated with the $m=3$ pinching. The scalar integral
for this pinching is not well behaved in this limit and so should not be
expected to combine with the other pinchings.
Therefore, we separate $\Ihat_5[1]$ according to the pole structure in $\e$ and
$N_5$.
We identify
the function $\Le{1}$ which is finite as both $N_5 \rightarrow 0$ and $\e \to
0$ and does not depend on the opposite-box pinching, $m=3$,
plus a combination of scalar box integrals containing
all the infrared poles and the remaining $N_5 \to 0$ singularities,
\begin{eqnarray}
\I_5[1] &=& \left(\sum_{i=1}^5 \alpha_3\alpha_i \kappa_i \right) \Le{1}
- \frac{1}{2N_5}
  \g_3 \I_4^{(3)}[1]
- \alpha_2\alpha_4\left( \I_4^{(2)}[1]+\I_4^{(4)}[1]\right )\nonumber \\
&&-\frac{1}{2}
\alpha_3 \left(
\alpha_1\I_4^{(1)}[1]-\alpha_2\I_4^{(2)}[1]-\alpha_4\I_4^{(4)}[1]
+\alpha_5\I_4^{(5)}[1]\right) ,
\end{eqnarray}
where
$\kappa_i = (1,-1,0,1,-1)$ and,
\begin{equation}
\Le{1} = -\frac{\alpha_1\alpha_2\alpha_4\alpha_5}{2N_5}
\sum_{m=1}^5  \kappa_m \Ihat_4^{(m)}[1] .
\end{equation}

Applying equation~(\ref{eq:Inxi}) and noting that $\I_5^{D=6}$ is both
infrared and ultraviolet finite, the integrals with one insertion are also
determined
in terms of box integrals in $D=4-2\e$ dimensions,
\begin{equation}
\Ihat_5[a_i] = -\frac{1}{2N_5}\sum_{m=1}^5
 \eta_{im} \Ihat_4^{(m)}[1] + \cal{O}(\e).
\end{equation}
Making the same separation as before,
\begin{equation}
\I_5[x_i] = \left(\alpha_3\alpha_i\kappa_i\right)
\Le{1}-\frac{1}{2N_5}
\sum_{m=1}^5  \alpha_i\alpha_m\left(\eta_{im}-\kappa_i\kappa_m\right)
\I_4^{(m)}[1] + \cal{O}(\e),
\end{equation}
where the bracket multiplying the scalar box integral is always proportional to
$N_5$.

Similarly,
using the usual formula for two Feynman parameters~(\ref{eq:Inxixj}) and
concentrating all of the $\Delta_5$ dependence into
$\chat_{i,j}$, where,
\begin{equation}
c_{i,j} = \alpha_i \alpha_j \chat_{i,j} =
\frac{\alpha_i\alpha_j}{2N_5}
\left(\eta_{ij} - \frac{\ghat_i\ghat_j}{\hat\Delta_5}\right),
\end{equation}
we find,
\begin{eqnarray}
\Ihat_5[a_ia_j] &=&
 \chat_{i,j}
\left( \Ihat_5^{D=6}[1] + \frac{1}{2N_5}
\sum_{m=1}^5 \ghat_m \Ihat_4^{D=6,(m)}[1] \right)\nonumber \\
&&
+ \frac{1}{4N_5^2} \sum_{m=1}^5 ( \eta_{im} \ghat_j - \eta_{ij}
 \ghat_m ) \Ihat_4^{D=6,(m)}[1] - \frac{1}{2N_5} \sum_{m=1}^5 \eta_{jm}
 \Ihat_4^{(m)}[a_i] \nonumber \\
&=&
- \chat_{i,j} \frac{\hat\Delta_5 \Ihat_5^{D=8}[1]}{N_5}
\nonumber \\
&&
+\frac{1}{2N_5}
\sum_{m=1}^5 \left(
\left( \frac{\ghat^{(m)}_j \eta_{im}
- \ghat_m \eta_{ij}^{(m)}}{\eta_{mm}} \right)
\Ihat_4^{D=6,(m)}[1] - \eta_{jm} \Ihat_4^{(m)}[a_i]\right).
\label{eq:pent-2ins}
\end{eqnarray}
To simplify the non-$\chat_{i,j}$ terms (and eliminate one power of $N_5$),
we have rewritten some of the variables appropriate to
the pentagon integral in terms of those appropriate to the box
pinchings, $\eta_{ij}^{(m)}$ and $\ghat_i^{(m)}$,
using the relations~\cite{BDK2},
\begin{displaymath}
\g_{i}^{(m)} = \frac{\eta_{mm}\g_{i}-\eta_{im}\g_{m}}{2N_5},\qquad
\eta_{ij}^{(m)} = \frac{\eta_{mm}\eta_{ij}-\eta_{im}\eta_{jm}}{2N_5} .
\end{displaymath}

Bern, Dixon and Kosower have shown~\cite{BDK1,BDK2} that $\I_5^6$ drops out of
the calculation of any gauge theory amplitudes by using the identity
\cite{EGY},
\begin{equation}
\sum_{i,j=1}^4 q_i^{\mu_i} q_j^{\mu_j} c_{i+1,j+1}
 = \frac{1}{2} g^{\mu_i\mu_j }_{[4]},
\label{eq:trick}
\end{equation}
where $g^{\mu_i \mu_j}_{[4]}$ represents the metric tensor in $D=4$.
Since,
\begin{eqnarray}
\I_5[\ell^{\mu_i}\ell^{\mu_j}] &=& \I_5[\P^{\mu_i}\P^{\mu_j}]
- \frac{1}{2}\I_5^{D=6}[1]g^{\mu_i\mu_j} \nonumber \\
&=& \sum_{i,j=1}^5 \I_5[x_{i+1}x_{j+1}]q_i^{\mu_i}q_j^{\mu_j}
- \frac{1}{2}\I_5^{D=6}[g^{\mu_i\mu_j}],
\end{eqnarray}
{\em all} terms proportional to $\chat_{i,j}$ in $\Ihat_5[a_ia_j]$
can be moved into the existing $g^{\mu_i\mu_j}$ piece.
Inspection of equation~(\ref{eq:pent-2ins}) indicates that
the correct term to reshuffle is $-\Delta_5 \Ihat_5^8[1]/N_5$ rather
than merely $\Ihat_5^{D=6}[1]$.
Retaining only the $c_{i,j}$ terms,
\begin{eqnarray}
g^{\mu_i\mu_j} \mbox{ terms} & = & -\frac{1}{2} \I_5^{D=6}[1]
 g^{\mu_i\mu_j}
 - \frac{\hat\Delta_5 \I_5^{D=8}[1]}{N_5}
 \sum_{i,j=1}^4 q_i^{\mu_i} q_j^{\mu_j} c_{i+1,j+1} \nonumber \\
&=&
-\frac{1}{2} \I_5^{D=6}[1]  g^{\mu_i\mu_j}  + \frac{1}{2}
 \left( \I_5^{D=6}[1] + \frac{1}{2N_5}
 \sum_{m=1}^5 \g_m \I_4^{D=6,(m)}[1] \right) g^{\mu_i\mu_j}_{[4]}\nonumber \\
&=&
\frac{1}{4N_5} \sum_{m=1}^5 \g_m \I_4^{D=6,(m)}[1]
 g^{\mu_i \mu_j} + {\cal O}(\e).
\end{eqnarray}
As expected the $\I_5^{D=6}[1]$ terms precisely cancel.
Here the finiteness of $\I_5^{D=6}[1]$ and $\I_5^{D=8}[1]$
has been used to ensure that this term
generates only ${\cal O}(\e)$ corrections when replacing
$g^{\mu_i\mu_j}_{[4]}$ with the dimensionally regularised $g^{\mu_i\mu_j}$.
The remaining piece
should not contain any kinematic singularities associated with lower-point Gram
determinants, in particular $N_5$, as it originates from the well-behaved
$\I_5^{D=8}$. This is indeed the case and we write,
\begin{equation}
\Le{2} = \frac{1}{2N_5} \sum_{m=1}^5 \g_m \I_4^{D=6,(m)}[1].
\end{equation}

The pentagon integrals with three insertions can be obtained by direct
differentiation of equation~(\ref{eq:pent-2ins}).
We find,
\begin{eqnarray}
\Ihat_5[a_ia_ja_k] &=&
\chat_{i,j} \left( \Ihat_5^{D=6}[a_k] + \frac{1}{6N_5}
 \sum_{m=1}^5 \left( \eta_{km} \Ihat_4^{D=6,(m)}[1] + 2\ghat_m
\Ihat_4^{D=6,(m)}[a_k]
 \right) \right)   \nonumber \\
&& +
\frac{1}{6N_5}
 \sum_{m=1}^5 \Biggl(
 2 \left( \frac{\ghat^{(m)}_j \eta_{im}  - \ghat_m \eta_{ij}^{(m)}}{\eta_{mm}}
 \right) \Ihat_4^{D=6,(m)}[a_k] \nonumber \\
&& - \left( \frac{\eta_{im}\eta_{jk}^{(m)} -
\eta_{ij}^{(m)}\eta_{km}}{\eta_{mm}}
 \right) \Ihat_4^{D=6,(m)}[1] -  \eta_{jm} \Ihat_4^{(m)}[a_ia_k] \Biggr)
+ \mbox{cyclic i,j,k}.
\label{eq:pent-3ins}
\end{eqnarray}
Since the $\chat_{i,j}$ term is obtained by differentiating
$c_{i,j}\I_5^{D=8}[1]$,
it must be finite as both $N_5 \to 0$ and $\e \to 0$ and we
introduce the finite function,
\begin{equation}
\Le{3k} = \frac{1}{6N_5} \sum_{m=1}^5 \left( \alpha_k\alpha_m\eta_{km}
\I_4^{D=6,(m)}[1]
 + 2\g_m \I_4^{D=6,(m)}[x_k] \right) .
\end{equation}
Recalling that,
\begin{eqnarray}
\I_5[\ell^{\mu_i}\ell^{\mu_j}\ell^{\mu_k}] &=&
\I_5[\P^{\mu_i}\P^{\mu_j}\P^{\mu_k}]
-\frac{1}{2}\I_5^{D=6}[\{ g\P \}^{\mu_i\mu_j\mu_k}]  \nonumber \\
&=& - \sum_{i,j,k=1}^5 \I_5[x_{i+1}x_{j+1}x_{k+1}]q_i^{\mu_i} q_j^{\mu_j}
q_k^{\mu_k}
\nonumber \\
&& +\frac{1}{2}\left\{
\sum_{k=1}^5 \I_5^{D=6}[x_{k+1}]g^{\mu_i\mu_j}q_k^{\mu_k}
+ \mbox{cyclic i,j,k} \right\},
\end{eqnarray}
and keeping only the $c_{i,j}$ terms in equation~(\ref{eq:pent-3ins})
we find,
\begin{eqnarray}
g^{\mu_i\mu_j}q_k^{\mu_k} \mbox{ terms} & = &  \frac{1}{2} \I_5^{D=6}[x_{k+1}]
 g^{\mu_i \mu_j}q_k^{\mu_k}
- \left( \I_5^{D=6}[x_{k+1}] + \Le{3k} \right)
 \sum_{i,j=1}^4  c_{i+1,j+1} q_i^{\mu_i} q_j^{\mu_j}q_k^{\mu_k} \nonumber \\
&=&
 \frac{1}{2} \I_5^{D=6}[x_{k+1}]  g^{\mu_i \mu_j}q_k^{\mu_k}
- \frac{1}{2}
 \left( \I_5^{D=6}[x_{k+1}] + \Le{3k} \right) g^{\mu_i \mu_j}_{[4]}q_k^{\mu_k}
\nonumber \\
&=&
-\frac{1}{2} \Le{3k}g^{\mu_i \mu_j}q_k^{\mu_k}  + {\cal O}(\e).
\end{eqnarray}
As in the previous case, the finiteness of the coefficient of
$g^{\mu_i \mu_j}_{[4]}$ has been used to promote it to the full
$D=4-2\e$ metric tensor.

The tensor integrals with four and five insertions may be obtained
by further differentiation, and the same trick used to
rewrite the $c_{i,j}$ terms as a contribution to the metric tensor structure.
In this way, all vestiges of the pentagon in $D=6-2\e$ and higher dimensions
can be removed, along with the inverse powers of $\Delta_5$.

\begin{figure}[t]\vspace{8cm}
\includegraphics{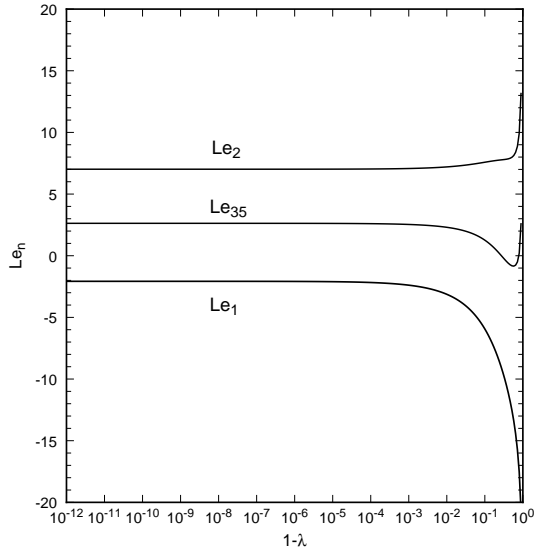}
\caption[]{The finite functions for the pentagon graph with
one external mass and evaluated in double precision Fortran
as a function of  $1-\lambda$. Because the functions only
contain single inverse powers of $(1-\lambda)$, no numerical problems are
evident.}
\label{fig:les}
\end{figure}

The functions introduced  in this section, $\Le{1}$, $\Le{2}$ and
$\Le{3i}$, only contain a single power
of the Gram determinant.
Consequently, they can be evaluated without numerical problems.
To illustrate this, we choose a representative phase space point,
$s_{1234}=1,~s_{123}=0.4,~s_{234}=0.3,~s_{13}=0.1$ and use
the variable $s_{23}$ to control $\lambda$.
The sixth variable $s_{24}$ is chosen to lie within the physical
region defined by the
pentagon Gram determinant, $\Delta_5 < 0$.
Fig.~\ref{fig:les} shows the functions $\Le{1}$ and $\Le{2}$
together with   $\Le{35}$.
In each case, we see that the $\lambda \to 1$ limit is smoothly approached
indicating that the function is intrinsically well behaved in that limit.

\section{Conclusions}

In this paper we have developed a new strategy for evaluating
one-loop tensor integrals. It avoids the usual problems
associated with the presence of Gram determinants.
Such Gram determinants arise when the tensor integrals are expressed
in terms of the physical momenta and generate false singularities at
the edges of phase space.
In addition to creating numerical instabilities,  they tend
to increase the size of the one-loop matrix elements.
Our approach is to construct groups of scalar integrals which are
well behaved in the limit of vanishing Gram determinant ($\Gn \to 0$),
and which can be evaluated
with arbitrary precision by making a Taylor series expansion
in $\Gn$.
In fact such combinations arise naturally by either differentiating with
respect to the external parameters - essentially yielding scalar
integrals with Feynman parameters in the numerator - or by developing
the scalar integral in $D=6-2\e$ or higher dimensions.
Evaluating these new integrals is straightforward - they are just
linear combinations of the known scalar integrals in $D=4$ or $D=4-2\e$.
As such, they  combine the dilogarithms, logarithms
and constants from different scalar integrals in an extremely
non-trivial way.
As a bonus other spurious kinematic singularities are also controlled
- they appear in the denominator of the finite functions, which are well
behaved in the singular limit.
Although the number of basic functions has increased, the number of
dilogarithm evaluations has not, since the functions are generated recursively.
Furthermore, because the Gram determinant singularities are not genuine,
by grouping integrals in this way, the expressions for one
loop integrals are compactified.

To illustrate the approach for specific integrals,
we have applied the method to 3-, 4- and 5-point integrals
where the internal masses have been set equal to zero.   These
tensor integrals are relevant for a range of QCD processes where
the quark and gluon masses
are negligible.
For more general processes with arbitrary internal masses and external
kinematics,
the relevant combinations of scalar
integrals can be obtained using
equations~(\ref{eq:d-change},\ref{eq:Inxi}--\ref{eq:Inxixjxkxl}).
As a by-product we have shown how all the Gram
determinants associated with pentagon graphs can be eliminated.

\section*{Acknowledgements}

We thank Walter Giele, Eran Yehudai, Bas Tausk and Keith Ellis
for collaboration in the earlier stages of this work.
EWNG thanks Alan Martin for useful discussions and for
suggestions concerning the manuscript.
JMC and DJM thank the UK Particle Physics and Astronomy Research Council
for the award of research studentships.

\appendix

\newpage
\section{Triangle integrals}
\setcounter{equation}{0}

In this appendix, we collect together explicit forms for the
triangle graphs which appear as building blocks in the
box graphs.
We fix the kinematics according to fig.~1, so that momenta
$p_1$ and $p_2$ are exiting, with $p_3$ determined by momentum
conservation.
There are two distinct cases, according to whether one or two of
the external masses are zero.

\subsection{The one-mass triangle}

Here we provide explicit results for the case $p_1^2 = p_2^2 = 0$.
There is an additional symmetry under the exchange $p_1 \leftrightarrow p_2$
and $x_1 \leftrightarrow x_3$.
Insertions involving $x_2$ can be eliminated using $x_1+x_2+x_3 = 1$.

The $D=4-2\e$ scalar and tensor integrals are given by,
\begin{eqnarray}
\I^{1m}_3[1] &=& \frac{c_\Gamma}{\epsilon^2} \frac{ (-s_{12})^{-\e}
}{s_{12}},\\
\I^{1m}_3[x_1] = \I^{1m}_3[x_3] = 2 \I^{1m}_3[x_1^2] = 2 \I^{1m}_3[x_3^2] &=&
- \frac{1}{s_{12}} \left( \frac{ (-s_{12})^{-\epsilon} }{\e} +2
\right)c_\Gamma,\\
\I^{1m}_3[x_1 x_3] = 3\I^{1m}_3[x_1^2 x_3] = 3\I^{1m}_3[x_1 x_3^2]
&=& \frac{1}{2s_{12}},\\
\I^{1m}_3[x_1^3] = \I^{1m}_3[x_3^3] &=&
 - \frac{1}{3s_{12}} \left( \frac{ (-s_{12})^{-\epsilon} }{\e} +
 \frac{13}{6} \right)c_\Gamma,
\end{eqnarray}
while, the necessary integrals in $D=6-2\e$ dimensions read,
\begin{eqnarray}
\I^{1m,D=6-2\e}_3[1] &=&
-\frac{1}{2} \left( \frac{ (-s_{12})^{-\epsilon} }{\e}
 +3 \right)c_\Gamma,\\
\I^{1m,D=6-2\e}_3[x_1] = \I^{1m,D=6-2\e}_3[x_3] &=&
 -\frac{1}{6}
\left( \frac{ (-s_{12})^{-\epsilon} }{\e} + \frac{8}{3} \right)c_\Gamma.
\end{eqnarray}

\subsection{The two-mass triangle}

When two of the external legs are massive,
$p_2^2=0$ but $s_{12},~p_1^2 \neq 0$, the divergent integrals read,
\begin{eqnarray}
\I^{2m}_3[1] &=& \frac{c_\Gamma}{\epsilon^2} \left( \frac{ (-s_{12})^{-\e}
 - (-p_1^2)^{-\e} }{s_{12}-p_1^2} \right),\\
\I^{2m}_3[x_3] &=& -2\Lc{2}^{2m}(p_1,p_2)
- \frac{c_\Gamma}{\e} \frac{ (-s_{12})^{-\e} }
 {(s_{12}-p_1^2)} - \frac{p_1^2}{(s_{12}-p_1^2)} \I^{2m}_3[1],\\
\I^{2m}_3[x_3^2] &=& -3\Lc{3}^{2m}(p_1,p_2) + \frac{1}{2(s_{12}-p_1^2)} -
 c_\Gamma\frac{ (-s_{12})^{-\e} }{\e} \frac{(s_{12}-3p_1^2)}{2(s_{12}-p_1^2)^2}
\nonumber \\
&&\qquad
 + \frac{p_1^4}{(s_{12}-p_1^2)^2} \I^{2m}_3[1],\\
\I^{2m}_3[x_3^3] & = & -\frac{11}{3} \Lc{4}^{2m}(p_1,p_2) +
\frac{(s_{12}-2p_1^2)}
 {2(s_{12}-p_1^2)^2} - \frac{p_1^6}{(s_{12}-p_1^2)^3} \I^{2m}_3[1] \nonumber
 \\ & & \qquad - \frac{ (-s_{12})^{-\e} }{\e}
 \left( \frac{1} {3(s_{12}-p_1^2)} - \frac{p_1^2}{2(s_{12}-p_1^2)^2}
 + \frac{p_1^4}{(s_{12}-p_1^2)^3} \right)c_\Gamma,
\end{eqnarray}
while,
\begin{eqnarray}
\I^{2m}_3[x_1] =2\I^{2m}_3[x_1^2] = 3\I^{2m}_3[x_1^3] &=&
\Lc{1}^{2m}(p_1,p_2),\\
\I^{2m}_3[x_1x_3] =3\I^{2m}_3[x_1^2x_3] &=& \frac{1}{2} \Lc{2}^{2m}(p_1,p_2),\\
\I^{2m}_3[x_1 x_3^2] &=& \frac{1}{3} \Lc{3}^{2m}(p_1,p_2).
\end{eqnarray}
The functions $\Lc{n}^{2m}(p_1,p_2)$ are defined by
equations~(\ref{eq:lcmassless},\ref{eq:lcmassless1}).

The $D=6-2\e$ dimension integrals read,
\begin{eqnarray}
\I^{2m,D=6-2\e}_3[1] &=& \frac{1}{2} \left( p_1^2 \Lc1^{2m}(p_1,p_2)
 - \left(\frac{ (-s_{12})^{-\epsilon} }{\e} + 3\right) c_\Gamma \right),\\
\I^{2m,D=6-2\e}_3[x_1] &=& \frac{1}{6} \left( p_1^2\Lc1^{2m}(p_1,p_2)
 - \left( \frac{ (-s_{12})^{-\epsilon} }{\e} + \frac{8}{3} \right)c_\Gamma
\right),\\
\I^{2m,D=6-2\e}_3[x_3] &=& \frac{1}{6} \left( p_1^2\Lc2^{2m}(p_1,p_2)
 - \left(\frac{ (-s_{12})^{-\epsilon} }{\e} + \frac{8}{3} \right)c_\Gamma
\right).
\end{eqnarray}

The corresponding integrals for the case $p_1^2 = 0$, $p_2^2 \neq 0$ can
be obtained by using the above formulae with
the substitutions,
\begin{equation}
p_1 \leftrightarrow p_2, \qquad x_1 \leftrightarrow x_3.
\end{equation}

The limit $p_1^2 \to 0$ may also be safely taken.  For example, using the fact
that in this limit $\Lc{n}^{2m}(p_1,p_2) \to 1/(n-1)/s_{12}$
and observing that all of the triangle loop integral contributions proportional
to $1/\e^2$ trivially drop out in eqs.~(A.8)-(A.10), we see that,
$$
\I_3^{2m}[x_3^n] \to \I_3^{1m}[x_3^n].
$$

\newpage
\section{Box integrals}
\setcounter{equation}{0}
Here we collect together explicit forms for the
some of the finite functions appearing in
the box integrals.
We fix the kinematics according to fig.~2, so that momenta
$p_1$, $p_2$ and $p_3$ are exiting, with $p_4$ determined by momentum
conservation.

\subsection{The adjacent two-mass box}

We first focus on the case where two of the adjacent external legs are
massive, $p_4^2 \neq 0$ and $p_1^2 \neq 0$.

For integrals with a single Feynman parameter in the numerator, we have,
\begin{equation}
\Ld{1}(p_1,p_2,p_3) = - \left( \Ld{1S}(p_1,p_2,p_3) + \Lc{0}(p_1,p_{23})
\right),
\end{equation}
where the box function  in $D=6$, $\Ld{1S}$ is given by
equation~(\ref{eq:Ld1sadj}).
When there are two Feynman parameters in the numerator, we eliminate $x_3$
using
$x_1 + x_2 + x_3 + x_4 = 1$ so there are only three relevant functions.
Explicitly, we find,
\begin{eqnarray}
\Ld{21}(p_1,p_2,p_3) &=& -\frac{2}{s_{12}} \left( 3\Ld{2S}(p_1,p_2,p_3)
 + \Lc{1S}(p_1,p_{23}) \right) - \Lc{1}(p_{23},p_1)  ,\\
\Ld{22}(p_1,p_2,p_3) & = & \frac{2(s_{123}-s_{12})}{s_{12}s_{23}}
 \left( 3\Ld{2S}(p_1,p_2,p_3) + \Lc{1S}(p_1,p_{23}) \right)
 \nonumber \\ & & \qquad - \frac{s_{12}}{s_{23}}\Lc{1}^{2m}(p_1,p_2)
 + \frac{s_{123}}{s_{23}} \Lc{1}(p_{23},p_1),\\
\Ld{24}(p_1,p_2,p_3) & = & \frac{2(p_1^2-s_{12})}{s_{12}s_{23}} \left(
 3\Ld{2S}(p_1,p_2,p_3) + \Lc{1S}(p_1,p_{23}) \right) \nonumber \\ & & \qquad -
 \frac{s_{12}}{s_{23}} \Lc{1}^{2m}(p_{12},p_3) + \frac{p_1^2}{s_{23}}
\Lc{1}(p_{23},p_1).
\end{eqnarray}
The all massive triangle integral function in $D=6-2\e$, $\Lc{1S}$ is given
by equation~(\ref{eq:Lc1s}), the box integral function in
$D=8-2\e$ is given in equation~(\ref{eq:Ld2sadj}) while the remaining
triangle functions are given in Appendix A and
equations~(\ref{eq:I3x3}--\ref{eq:I3x3x3x3}).
The functions for adjacent box integrals with three insertions of Feynman
parameters contain the box in $D=10-2\e$ (\ref{eq:Ld3sadj})
and triangle in $D=8-2\e$ (\ref{eq:Lc2s}).
All integrals can be obtained in terms of the following four functions,
\begin{eqnarray}
\Ld{311}(p_1,p_2,p_3) & = & - \Biggl( \frac{12}{s_{12}^2} \left(
 5\Ld{3S}(p_1,p_2,p_3) + \Lc{2S}(p_1,p_{23}) \right)
\nonumber \\ & &
+ \frac{(s_{12}+p_1^2)}{s_{12}}\Lc{3}(p_{23},p_1)
+ \frac{s_{23}}{s_{12}} \Lc{2}(p_1,p_{23}) \Biggr),
\\
\Ld{314}(p_1,p_2,p_3) & = & \frac{12(p_1^2-s_{12})}{s_{12}^2s_{23}}
 \left( 5\Ld{3S}(p_1,p_2,p_3) + \Lc{2S}(p_1,p_{23}) \right)
\nonumber \\ &&
+ \frac{p_1^4}{s_{12}s_{23}} \Lc{3}(p_{23},p_1)
- \frac{(s_{12}-p_1^2)}{s_{12}} \Lc{2}(p_1,p_{23})
\nonumber \\ & &
 -\frac{s_{12}}{2s_{23}}  \Lc{1}^{2m}(p_{12},p_3),
\\
\Ld{322}(p_1,p_2,p_3) & = & - \Biggl(
\frac{12(s_{123}-s_{12})^2}{s_{12}^2s_{23}^2}
 \left( 5\Ld{3S}(p_1,p_2,p_3) + \Lc{2S}(p_1,p_{23}) \right)
\nonumber \\ & &
+ \frac{s_{123}(s_{12}s_{123}+p_1^2s_{123}-2p_1^2s_{12})}{s_{12}s_{23}^2}
              \Lc{3}(p_{23},p_1)
- \frac{s_{12}}{2s_{23}} \Lc{2}^{2m}(p_1,p_2)
\nonumber \\ & &
+ \frac{s_{12}(s_{12}-s_{123})}{2s_{23}^2}\Lc{1}^{2m}(p_{12},p_3)
+ \frac{s_{123}(s_{123}-2s_{12})}{s_{12}s_{23}} \Lc{2}(p_1,p_{23})
\nonumber \\ & &
+ \frac{(p_1^2s_{12}+s_{12}s_{23}-s_{12}s_{123})}{2s_{23}^2}
         \Lc{1}^{2m}(p_1,p_2) \Biggr),
\\
\Ld{344}(p_1,p_2,p_3) & = & - \Biggl(
\frac{12(s_{12}-p_1^2)^2}{s_{12}^2s_{23}^2}
 \left( 5\Ld{3S}(p_1,p_2,p_3) + \Lc{2S}(p_1,p_{23}) \right)
\nonumber \\ & &
 + \frac{p_1^2(p_1^2-2s_{12})}{s_{12}s_{23}} \Lc{2}(p_1,p_{23})
 + \frac{p_1^4(p_1^2-s_{12})}{s_{12}s_{23}^2} \Lc{3}(p_{23},p_1)
\nonumber \\ & &
 + \frac{s_{12}(s_{12}-p_1^2)}{2s_{23}^2} \Lc{1}^{2m}(p_{12},p_3)
 + \frac{s_{12}}{2s_{23}} \Lc{2}^{2m}(p_{12},p_3) \Biggr).
\end{eqnarray}

\subsection{The one-mass box}

For this kinematic configuration, only $p_4^2 \neq 0$ and there is a
`flip' symmetry,
so that functions related to the parameter $x_4$ are obtained from
those
related to $x_1$, by $p_1 \longleftrightarrow p_3$.
The box integrals in higher dimension are given by
equations~(\ref{eq:Ld1s1m}--\ref{eq:Ld4s1m}),
and,
\begin{equation}
\Ld{1}^{1m}(p_1,p_2,p_3) = -\Ld{1S}^{1m}(p_1,p_2,p_3).
\end{equation}
For two and three insertions, we find,
\begin{eqnarray}
\Ld{21}^{1m}(p_1,p_2,p_3) &=& -\frac{1}{s_{12}}
\left( 6 \Ld{2S}^{1m}(p_1,p_2,p_3) + s_{23} \Lc{1}^{2m}(p_{23},p_1) \right),\\
\Ld{22}^{1m}(p_1,p_2,p_3) &=&
\Ld{1}^{1m}(p_1,p_2,p_3)-\Ld{21}^{1m}(p_1,p_2,p_3),
\end{eqnarray}
and,
\begin{eqnarray}
\Ld{311}^{1m}(p_1,p_2,p_3) &=& -\frac{1}{s_{12}^2}
\Biggl( 60 \Ld{3S}^{1m}(p_1,p_2,p_3)
 + \frac{1}{2} s_{23}^2 \Lc{1}^{2m}(p_{23},p_1)
\nonumber \\ & &
+ \frac{1}{2} s_{12}s_{23} \Lc{2}^{2m}(p_{23},p_1)
\Biggr),\\
\Ld{314}^{1m}(p_1,p_2,p_3)&=& -\frac{1}{s_{12}s_{23}}
\Biggl( 60 \Ld{3S}^{1m}(p_1,p_2,p_3)
 + \frac{1}{2} s_{23}^2 \Lc{1}^{2m}(p_{23},p_1)
\nonumber \\ & &
+ \frac{1}{2} s_{12}^2 \Lc{1}^{2m}(p_{12},p_3)
\Biggr),\\
\Ld{322}^{1m}(p_1,p_2,p_3) & = &
\Ld{311}^{1m}(p_1,p_2,p_3)+\Ld{1}^{1m}(p_1,p_2,p_3)-2\Ld{21}^{1m}(p_1,p_2,p_3).
\end{eqnarray}

\newpage
\section{Limits}
\setcounter{equation}{0}

In this section, we collect together suitable expansions of
the various functions presented in this paper in the limit
$\Delta_n \to 0$.
These expressions represent the leading term in the expansion of the
functions as a Taylor series in $\Delta_n$
and should be
evaluated for $\Delta_n <  \delta$, where $\delta$ can be
determined numerically.
Typically, $\delta \sim 10^{-4} \Delta_n^{{\rm max}}$ where
$\Delta_n^{{\rm max}}$ is the largest value the Gram determinant
can achieve.
In general, for a given numerical precision, $acc$
the numerical problems occur
when $\Delta \sim (acc)^{1/N}$ where $N$ is the
number of Gram determinants in the denominator of the function.

\subsection{The three-mass triangle}

In the limit that $\Delta_3 \to 0$, we have,
\begin{eqnarray}
\Lc{0}(p_1,p_2) &\to & \frac{1}{2p_1^2p_2^2s_{12}}
\left(p_1^2(s_{12}+p_2^2-p_1^2)\log\left(\frac{s_{12}}{p_1^2}\right)
+
 p_2^2(s_{12}+p_1^2-p_2^2)\log\left(\frac{s_{12}}{p_2^2}\right) \right),
\nonumber \\
&&\\
\Lc{1S}(p_1,p_2) &\to & \frac{1}{12p_1^2p_2^2s_{12}}
\Biggl(p_1^4(s_{12}+p_2^2-p_1^2)\log\left(\frac{s_{12}}{p_1^2}\right)
+
 p_2^4(s_{12}+p_1^2-p_2^2)\log\left(\frac{s_{12}}{p_2^2}\right)
\nonumber \\ &&\qquad
+2p_1^2p_2^2s_{12} \Biggr),
\\
\Lc{2S}(p_1,p_2) &\to & \frac{1}{120p_1^2p_2^2s_{12}}
\Biggl(p_1^6(s_{12}+p_2^2-p_1^2)\log\left(\frac{s_{12}}{p_1^2}\right)
+
 p_2^6(s_{12}+p_1^2-p_2^2)\log\left(\frac{s_{12}}{p_2^2}\right)
\nonumber \\ &&\qquad
+\frac{p_1^2p_2^2s_{12}}{2}(s_{12}+p_1^2+p_2^2) \Biggr),
\\
\Lc{3S}(p_1,p_2) &\to & \frac{1}{1680p_1^2p_2^2s_{12}}
\Biggl
(p_1^8(s_{12}+p_2^2-p_1^2)\log\left(\frac{s_{12}}{p_1^2}\right)
+
 p_2^8(s_{12}+p_1^2-p_2^2)\log\left(\frac{s_{12}}{p_2^2}\right)
\nonumber \\ &&\qquad
+\frac{2p_1^2p_2^2s_{12}}{9}
   (s_{12}^2+p_1^4+p_2^4+s_{12}p_1^2+s_{12}p_2^2+p_1^2p_2^2) \Biggr),
\\
\Lc{1}(p_1,p_2) &\to &
\frac{1}{3p_1^2(s_{12}+p_2^2-p_1^2)}
\left((s_{12}-p_2^2)\log\left(\frac{s_{12}}{p_2^2}\right)
+2p_1^2-p_1^4\Lc{0}(p_1,p_2)\right),
\\
\Lc{2}(p_1,p_2) &\to &
\frac{1}{6p_2^2(s_{12}+p_1^2-p_2^2)}
\Biggl(-4p_2^2(s_{12}+p_1^2-p_2^2) \Lc{3}(p_1,p_2)
\nonumber \\ &&\qquad
+s_{12}-p_2^2
+3p_1^2p_2^2\Lc{1}(p_1,p_2)-p_1^4\Lc{1}(p_2,p_1)
+p_2^2\log\left(\frac{s_{12}}{p_2^2}\right) \Biggr ),
\\
\Lc{3}(p_1,p_2) &\to &
\frac{1}{5p_1^2(s_{12}+p_2^2-p_1^2)}
\left((s_{12}-p_2^2)\log\left(\frac{s_{12}}{p_2^2}\right)
+p_1^2-2p_1^4\Lc{1}(p_1,p_2)\right),
\\
\Lc{4}(p_1,p_2) &\to &
\frac{1}{15p_2^2(s_{12}+p_1^2-p_2^2)}
\Biggl(-6p_2^2(s_{12}+p_1^2-p_2^2) \Lc{5}(p_1,p_2)
\nonumber \\ &&\qquad
+s_{12}-p_2^2
+5p_1^2p_2^2\Lc{3}(p_1,p_2)-4p_1^4\Lc{2}(p_1,p_2)
+p_2^2\log\left(\frac{s_{12}}{p_2^2}\right) \Biggr ),
\\
\Lc{5}(p_1,p_2) &\to &
\frac{1}{7p_1^2(s_{12}+p_2^2-p_1^2)}
\left((s_{12}-p_2^2)\log\left(\frac{s_{12}}{p_2^2}\right)
+\frac{2p_1^2}{3}-3p_1^4\Lc{3}(p_1,p_2)\right).
\end{eqnarray}

\subsection{The adjacent two-mass box}

In the limit $\Delta_4 \to 0$, we have,
\begin{eqnarray}
\Ld{0}(p_1,p_2,p_3) & \rightarrow &
 -\frac{1}{2} \left( s_{123}+p_1^2-s_{23}-\frac{2p_1^2s_{123}}
 {s_{12}} \right) \Lc{0}(p_1,p_{23}), \\
\Ld{1S}(p_1,p_2,p_3) & \rightarrow &
 \frac{-1}{s_{12}s_{23}} \left( \left( s_{123}+p_1^2-s_{23}-
 \frac{2p_1^2s_{123}} {s_{12}} \right) \Lc{1S}(p_1,p_{23})
 \right.
\nonumber \\
& & \qquad \left.
+ \frac{s_{23}}{2}
\log \left(\frac{s_{123}} {s_{23}} \right) +
s_{12} \log \left( \frac{s_{123}}{s_{12}} \right)
- \frac{p_1^2}{2}
\log \left( \frac{s_{123}}{p_1^2} \right) \right), \\
\Ld{2S}(p_1,p_2,p_3) & \rightarrow &
 \frac{-1}{12s_{12}s_{23}} \left( \left( s_{123}+p_1^2-s_{23}-
 \frac{2p_1^2s_{123}} {s_{12}} \right) \Lc{2S}(p_1,p_{23})
 \right. \nonumber \\ & & \qquad \left. + \frac{s_{23}^2}{2} \log
 \left( \frac{s_{123}} {s_{23}} \right) + s_{12}^2 \log
 \left( \frac{s_{123}}{s_{12}} \right) - \frac{p_1^4}{2}
\log \left(
 \frac{s_{123}}{p_1^2} \right) + s_{12}s_{23} \right), \nonumber \\
\\
\Ld{3S}(p_1,p_2,p_3) & \rightarrow &
 \frac{-1}{180s_{12}s_{23}} \left( \left( s_{123}+p_1^2-s_{23}-
 \frac{2p_1^2s_{123}} {s_{12}} \right) \Lc{3S}(p_1,p_{23})
 \right.
\nonumber \\
& & \qquad \left. + \frac{s_{23}^3}{2}
\log
 \left(  \frac{s_{123}} {s_{23}} \right) + s_{12}^3 \log
 \left(  \frac{s_{123}}{s_{12}} \right)
- \frac{p_1^6}{2} \log \left(\frac{s_{123}}{p_1^2} \right) \right.
\nonumber \\
& & \left. \qquad
+ \frac{1}{4}s_{12}s_{23}(s_{123}+s_{12}+s_{23}+p_1^2) \right),
\\
\Ld{4S}(p_1,p_2,p_3) & \rightarrow &
 \frac{-1}{840s_{12}s_{23}} \left( \left( s_{123}+p_1^2-s_{23}-
 \frac{2p_1^2s_{123}} {s_{12}} \right) \Lc{4S}(p_1,p_{23})
 \right.
\nonumber \\
& & \qquad \left. + \frac{s_{23}^4}{2} \log
 \left(  \frac{s_{123}} {s_{23}} \right) + s_{12}^4 \log
 \left(  \frac{s_{123}}{s_{12}} \right)
- \frac{p_1^8}{2} \log \left(\frac{s_{123}}{p_1^2} \right) \right.
\nonumber \\ & &
\left. \qquad
+ \frac{1}{9}s_{12}s_{23}(s_{123}^2+s_{12}^2+s_{23}^2+p_1^4
 +s_{123}s_{12}+s_{123}s_{23} \nonumber \right. \\ & & \qquad \left.
 +s_{123}p_1^2+s_{12}p_1^2+s_{23}p_1^2+\frac{s_{12}s_{23}}{2})  \right).
\end{eqnarray}
In this last equation, we have used the finite part of the three mass
triangle graph in $D=12-2\e$.  For $p_1^2,~p_2^2 \neq 0$,
\begin{eqnarray}
\Lc{4S}(p_1,p_2) &=&
\frac{1}{8\Delta_3}
\Biggl (2p_1^2p_2^2s_{12}\Lc{3S}(p_1,p_2)
-\frac{1}{840}\Biggl (p_1^8(s_{12}+p_2^2-p_1^2)
\log\left(\frac{s_{12}}{p_1^2}\right)
\nonumber \\ &&\qquad
+p_2^8(s_{12}+p_1^2-p_2^2)
\log\left(\frac{s_{12}}{p_2^2}\right)
\nonumber \\ &&\qquad
+\frac{2p_1^2p_2^2s_{12}}{9}
(p_1^4+p_2^4+s_{12}^2+p_1^2s_{12}+p_2^2s_{12}+p_1^2p_2^2)\Biggr)\Biggr).
\label{eq:Lc4s}
\end{eqnarray}

\subsection{The one-mass box}

Finally, in the limit $\Delta_4^{1m} \to 0$, we have,
\begin{eqnarray}
\Ld{0}^{1m} (p_1,p_2,p_3) & \rightarrow & 0, \\
\Ld{1S}^{1m} (p_1,p_2,p_3) & \rightarrow &
 \frac{-1}{s_{12}s_{23}} \left( s_{23} \log \left( \frac{s_{123}}
 {s_{23}} \right) + s_{12} \log \left( \frac{s_{123}}{s_{12}} \right)
 \right), \\
\Ld{2S}^{1m} (p_1,p_2,p_3) & \rightarrow &
 \frac{-1}{12s_{12}s_{23}} \left( s_{23}^2 \log \left( \frac{s_{123}}
 {s_{23}} \right) + s_{12}^2 \log \left( \frac{s_{123}}{s_{12}}
\right) +s_{12}s_{23}  \right), \\
\Ld{3S}^{1m} (p_1,p_2,p_3) & \rightarrow &
 \frac{-1}{180s_{12}s_{23}} \left( s_{23}^3 \log \left( \frac{s_{123}}
 {s_{23}} \right) + s_{12}^3 \log \left( \frac{s_{123}}{s_{12}}
\right) \right. \nonumber \\ & & \qquad \left.
 + \frac{1}{4}s_{12}s_{23}(s_{123}+s_{12}+s_{23}) \right), \\
\Ld{4S}^{1m} (p_1,p_2,p_3) & \rightarrow &
 \frac{-1}{840s_{12}s_{23}} \left( s_{23}^4 \log \left( \frac{s_{123}}
 {s_{23}} \right) + s_{12}^4 \log \left( \frac{s_{123}}{s_{12}}
\right) \right. \nonumber \\ & & \qquad \left.
 + \frac{1}{9}s_{12}s_{23} \left( s_{123}^2+s_{12}^2+s_{23}^2
   +s_{123}s_{12}+s_{123}s_{23}+\frac{s_{12}s_{23}}{2} \right)  \right)
.\nonumber \\
\end{eqnarray}

\end{document}